\documentclass[twocolumn]{aastex631}
\usepackage{amsmath}
\usepackage{color}
\usepackage{float}

\newcommand{\imcom}{{\sc Imcom}}
\newcommand{\Pyimcom}{{\sc PyImcom}}
\newcommand{\RomanT}{\textit{Roman}}

\usepackage[T2A,T1]{fontenc}
\DeclareSymbolFont{cyrillic}{T2A}{cmr}{m}{n}
\DeclareMathSymbol{\Sha}{\mathalpha}{cyrillic}{216}

\begin{document}

\title{Charge diffusion and modulation transfer function in a \textit{Nancy Grace Roman Space Telescope} detector}

\correspondingauthor{Emily Macbeth}
\email{macbeth@arizona.edu}

\author[0009-0002-7190-9775]{Emily Macbeth}
\affiliation{Center for Cosmology and AstroParticle Physics (CCAPP), The Ohio State University, 191 West Woodruff Ave, Columbus, OH 43210, USA}
\affiliation{Department of Physics, The Ohio State University, 191 West Woodruff Ave, Columbus, OH 43210, USA}
\affiliation{Department of Astronomy, The Ohio State University, 140 West 18th Avenue, Columbus, OH 43210, USA}
\affiliation{Department of Astronomy and Steward Observatory, University of Arizona, 933 North Cherry Avenue, Tucson, AZ 85721, USA}

\author[0000-0002-6111-6061]{Katherine Laliotis}
\affiliation{Center for Cosmology and AstroParticle Physics (CCAPP), The Ohio State University, 191 West Woodruff Ave, Columbus, OH 43210, USA}
\affiliation{Department of Physics, The Ohio State University, 191 West Woodruff Ave, Columbus, OH 43210, USA}
\author{Christopher M. Hirata}
\affiliation{Center for Cosmology and AstroParticle Physics (CCAPP), The Ohio State University, 191 West Woodruff Ave, Columbus, OH 43210, USA}
\affiliation{Department of Physics, The Ohio State University, 191 West Woodruff Ave, Columbus, OH 43210, USA}
\affiliation{Department of Astronomy, The Ohio State University, 140 West 18th Avenue, Columbus, OH 43210, USA}

\author{Christopher Merchant}
\affiliation{NASA Goddard Space Flight Center, Detector Systems Branch, 8800 Greenbelt Rd, Greenbelt, Maryland 20771, USA}

\collaboration{4}{for the Roman HLIS Cosmology Project Infrastructure Team}

\begin{abstract}
The Nancy Grace Roman Space Telescope (\RomanT) is an observatory motivated by the search to understand dark energy, exoplanets, and general astrophysics. \RomanT \, will bring unprecedented amounts of precision to weak gravitational lensing measurements, which necessitates an improved understanding of instrumental signatures in star and galaxy images. One feature is the modulation transfer function (MTF), which includes contributions from charge diffusion in \RomanT's infrared detector arrays. As part of the detector characterization effort, a detector from the flight lots (but ultimately not selected for flight) was illuminated with a laser speckle pattern. We present an analysis of the laser speckle data, including MTF measurements in several wavelengths. We fit several models for the charge diffusion profile, including: (i) a Gaussian profile; (ii) a hyperbolic secant (sech) profile; and (iii) a general drift-diffusion model that includes the Gaussian and sech as limiting cases. We find that the sech model produces an acceptable fit with no need for the additional parameter and is strongly preferred over the Gaussian. The standard deviation of the sech profile is $0.3279^{+0.0043}_{-0.0042}$(stat)$\pm0.0093$(sys) pixels, with the systematic error dominated by non-linearities. We find no detectable wavelength dependence over the range from 850--2000 nm. The model informs survey strategy for weak lensing measurements and has been included in simulations used to develop the data processing pipelines for the \RomanT \,mission.
\end{abstract}


\section{Introduction} \label{sec:intro}

One of the most important discoveries in modern cosmology was the accelerating expansion of the Universe \citep{1998AJ....116.1009R, 1999ApJ...517..565P}. Nearly three decades later, ever more precise measurements of cosmic expansion continue to challenge the current cosmological model \citep{2025arXiv250314739D, 2025arXiv250314738D}. Characterizing the kinematics of the expansion, testing theories of gravity by mapping the large-scale structure of the Universe, and exploring the growth of matter clustering over time are major goals for the current and upcoming cosmological surveys.

A leading method of mapping the large-scale structure and matter in the Universe is weak gravitational lensing \citep{Weinberg2013, Mandelbaum2018}. Weak gravitational lensing (WL) is the deflection of light rays from distant objects by matter along the line of sight, characterized by slight modifications of the background objects' properties, such as their position, size, brightness, and shape \citep{Mandelbaum2018}. WL is used to determine the distribution of matter in the universe and the growth of cosmic structure over time \citep{Wittman2000}. This, in turn, facilitates the ability to constrain dark energy properties \citep{Weinberg2013}. As WL signals are extremely small ($\sim$1$\%$ distortions), precise measurements and small statistical errors are necessary.

The next generation of weak lensing surveys are designed to significantly improve precision and accuracy of measurements, including the \textit{Euclid} \citep{Laureijs2011, 2025A&A...697A...1E} mission, the Vera C. Rubin Observatory Legacy Survey of Space and Time (LSST) \citep{Ivezic2019}, and the
\textit{Nancy Grace Roman Space Telescope} (\RomanT) High Latitude Wide Area Survey (HLWAS) \citep{Spergel2015, 2019BAAS...51c.341D}. These surveys will feature crucial advances in precision measurements, requiring a new level of control of systematic uncertainties in weak lensing measurements. The combined efforts of these surveys will result in an unprecedented accuracy of WL measurements. 

\RomanT \, and \textit{Euclid} are space-based observatories, allowing for higher angular resolution than ground-based observatories, such as Rubin. However, the high angular resolution of the telescopes introduces further complications with the pixel scale. For a circular aperture, the point-spread function (PSF) of a diffraction-limited telescope has an angular width characterized by $\lambda/D$, where $\lambda$ is the wavelength of the observed light and $D$ is the diameter of the aperture. If the pixel scale is larger than $\lambda/(2D)$, then an image is considered \textit{undersampled} by the Nyquist standards. While one solution to undersampled images is to reduce the pixel size, this would result in much longer survey times and a smaller field-of-view. To preserve efficiency in survey time and a larger field-of-view, \RomanT \, and \textit{Euclid} both accept the undersampled images. When the image is undersampled, the contribution of the pixel response to the PSF becomes more important, and its band limit can affect which dithering patterns allow the image to be interpolated \citep{1999PASP..111..227L}.

An accurate model of the effective PSF can be determined through a convolution of several components. These components are a combination of atmospheric, optical, and intrinsic \textit{Roman} detector effects which are studied and modeled thoroughly. Some detector effects, like charge diffusion, can be convolved with the PSF, while other nonlinear effects have to be corrected for through different regimes. By determining the charge diffusion in the \textit{Roman} detectors, the effective PSF can be properly modeled and applied to the extensive science goals that require precision measurements. Producing an accurate model of the PSF is crucial for many science objectives of the \textit{Roman} mission. For weak lensing analysis, the linear algebra image combination algorithm \imcom \, \citep{Rowe2011} and a newer implementation \Pyimcom \, \citep{Cao2025} minimizes both the noise in output images and the overall deviation from a "target" PSF. \Pyimcom \, must return an accurate model of the PSF so its effects can be properly corrected for in data analysis, reducing systematic error. An accurate model of the effective PSF is also necessary for Type Ia Supernovae (SN Ia) cosmology. Measurements from the High-Latitude Time-Domain survey \citep{Rose2021} will depend on an effective PSF to properly classify SN Ia with high precision. The ability to model a realistic PSF is also needed for the Galatic Bulge Time-Domain Survey \citep{Penny_2019} that will use gravitational microlensing to conduct a census of exoplanetary systems.

The \textit{Roman} detectors are mercury cadmium telluride (Hg$_{1-x}$Cd$_x$Te) photodiode arrays \citep{Rogalski2005}.
{\slshape Roman} uses the next-generation 4k$\times$4k detector arrays with 10 $\mu$m pitch (H4RG-10). Changes such as the smaller pixel size (versus 18 $\mu$m in the earlier H2RG detectors used on the {\slshape James Webb Space Telescope}) potentially introduce new behaviors that differ from arrays flown on previous space telescopes.

This paper focuses on the contribution of charge diffusion to the PSF. In the HgCdTe detectors, a grown-in electric field $E_z$ is present that will drive liberated charges (in this case, holes) down towards the $p-n$ junctions of the diodes \citep{Mosby2020}. In addition to the drift velocity given by the grown-in field, the holes are also undergoing diffusion in all 3 dimensions. The probability distribution of the final location where the hole is collected --- and therefore the contribution of charge diffusion to the PSF --- is determined by a combination of the drift and diffusion.

We quantify the result of charge diffusion in terms of its contribution to the modulation transfer function (MTF), the absolute value of the Fourier transform of the PSF. The MTF of a system describes its spatial frequency response to a sine wave fluctuation in the sky image and is a combination of several effects that multiply: the ideal pixel response (a top hat function), charge diffusion, and the inter-pixel capacitance (IPC). The MTF is measured through well-defined methods, allowing us to fit well-known special cases and a \textit{Roman}-specific model to determine the extent of charge diffusion present in the detectors. By determining the contribution of charge diffusion to the MTF, and thus to image degradation, we can make informed decisions for survey strategy and understand how charge diffusion behaves in the new H4RG-10 detectors. 

This paper is organized as follows. We briefly describe some conventions in Sec.~\ref{sec:conventions}. The data used are described in Sec.~\ref{sec:experiment}. We describe the extraction of the MTF from the data in Sec. \ref{sec: MTF}. We present fits to charge diffusion models in Sec.~\ref{sec:modelfit}, and discuss the implications for {\slshape Roman} in Sec.~\ref{sec:discussion}.

We have placed several technical results in the appendices. We provide a derivation of the relation of the speckle autocorrelation function to the MTF in Appendix~\ref{appendix_A}. Our solution of the drift-diffusion equation (which reduces to the commonly-used Gaussian and hyperbolic secant functions in the respective limits) is provided in Appendix~\ref{appendix_B}. Some results from alternative analyses are placed in Appendix~\ref{appendix_C}.

\section{Conventions}
\label{sec:conventions}

For continuous functions $f(x,y)$, we define a Fourier transform by
\begin{align}
\tilde f(u,v)
  &= \int_{-\infty}^\infty f(x,y)\,e^{-2\pi i(ux+vy)}\,dx\,dy 
  \nonumber\\[6pt]
&\!\!\!\!\!\!\!\!\!\!\!\!\!\!\leftrightarrow
  f(x,y) = \int_{-\infty}^\infty \tilde f(u,v)\,e^{2\pi i(ux+vy)}\,du\,dv
\end{align}

For discretely sampled data, $f_{m_x,m_y}$ on an $N_x\times N_y$ grid, we compute a discrete Fourier transform by
\begin{align}
f_{m_x,m_y}
  &= \sum_{j_x=0}^{N_x-1} \sum_{j_y=0}^{N_y-1}
  \tilde f_{j_x,j_y}\,
   e^{2\pi i\!\left(\frac{j_x m_x}{N_x} + \frac{j_y m_y}{N_y}\right)}
  \nonumber\\[6pt]
&\!\!\!\!\!\!\!\!\!\!\!\!\!\!\leftrightarrow~
  \tilde f_{j_x,j_y} = \frac{1}{N_x N_y}
     \sum_{m_x=0}^{N_x-1} \sum_{m_y=0}^{N_y-1}
     f_{m_x,m_y}\,
     e^{-2\pi i\!\left(\frac{j_x m_x}{N_x} + \frac{j_y m_y}{N_y}\right)}
\label{eq:DFT}
\end{align}

The Fourier frequency may be indexed by its integer $(j_x,j_y)$. Alternatively, we will sometimes talk in terms of ``cycles per pixel'' -- that is, a given Fourier mode has spatial frequency $j_x/N_x$ cycles per pixel in the horizontal direction and $j_y/N_y$ cycles per pixel in the vertical direction.

We define the sinc function to include a $\pi$ so that the zeroes are at $\pm 1,\pm 2,\pm 3,...$:
\begin{equation}
{\rm sinc}\,\zeta \equiv \frac{\sin(\pi \zeta)}{\pi \zeta};
\end{equation}
this follows the same convention used in numpy.
We also define the sampling function or $\Sha$ (``Sha'')-function:
\begin{equation}
\Sha(\zeta)\equiv \sum_{q=-\infty}^\infty \delta(\zeta-q)
\end{equation}

where $\delta$ is the Dirac delta function. 

\section{Experimental Setup and Data} \label{sec:experiment}

\begin{figure}[]
\centering
\includegraphics[width=0.45\textwidth]{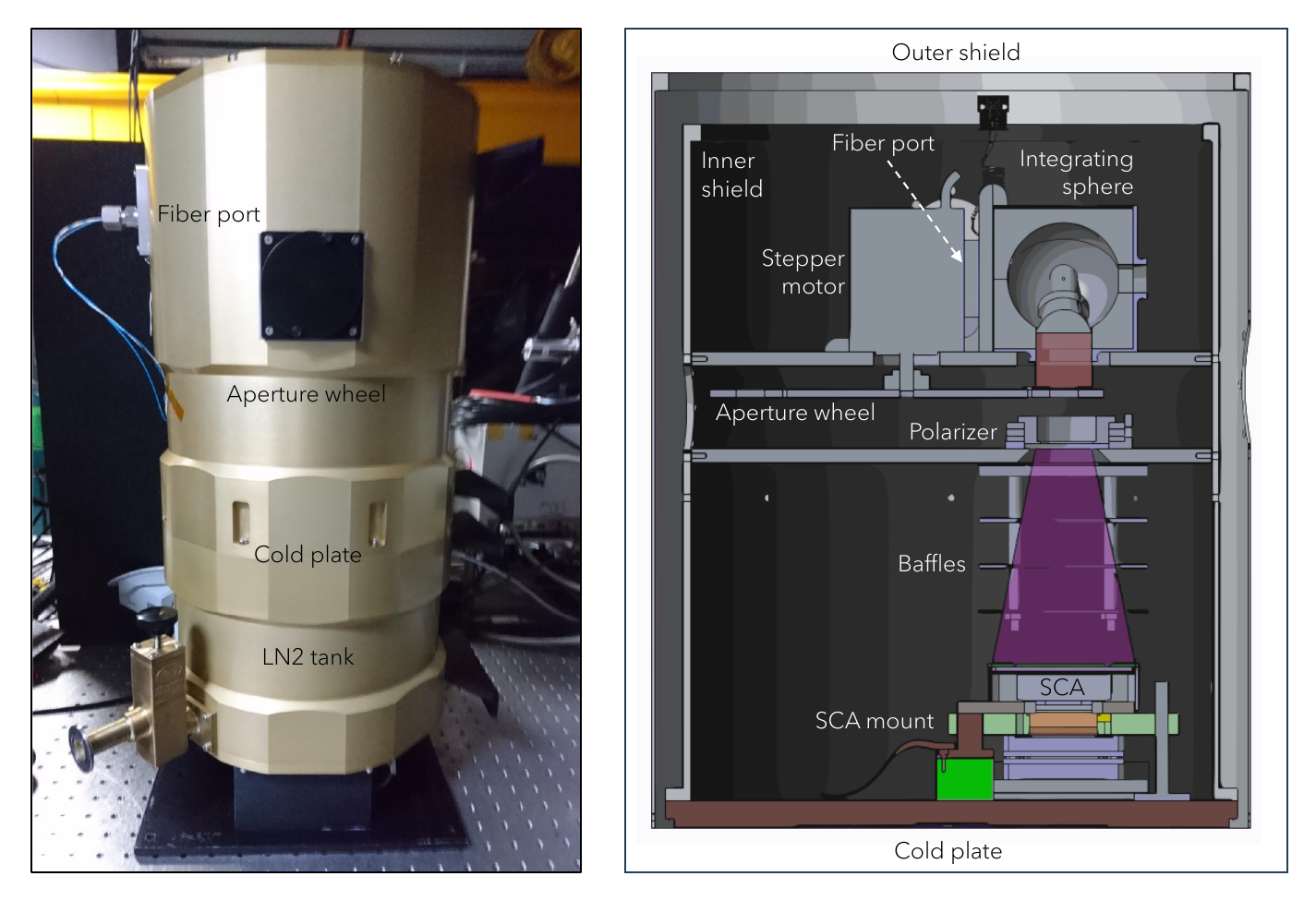}
\caption{\label{fig:setup}The setup for this experiment, at the Detector Characterization Laboratory at NASA Goddard Space Flight Center.}
\end{figure}

The MTF is measured utilizing laser-speckle patterns. This method is simple as it does not require critical alignment of the system or an optical lens, and it allows for the entire array of the detector to be tested \citep{Sensiper1993, Pozo2005}. The laser-speckle experiment uses double-slit apertures because they preserve more precise measurements at higher spatial frequencies. A single-slit aperture contains all spatial frequencies, which can lead to aliasing when measurements are above the Nyquist frequency, where the Fourier modes of high frequencies are indistinguishable from low frequencies. Using the double-slit, for which this issue is not present due to discrete peaks in Fourier space because of slit separation, avoids further complications with measurements.  

\begin{table}[H]
    \begin{tabular}{|c|c c|}
        \hline
        Aperture & Slit spacing & Slit width \\ & $d$ (mm) & $w$ (mm) \\
        \hline
        AP1 & 0.4 & 0.05 \\ 
        AP2 & 0.8 & 0.05\\
        AP3 & 1.6 & 0.2\\
        AP4 & 2.0 & 0.2\\
        AP5 & 3.0 & 0.2\\
        AP6 & 4.0 & 0.2\\
        AP7 & 5.0 & 0.3\\
        AP8 & 6.0 & 0.3\\
        AP9 & 7.0 & 0.3\\
        AP10 & 9.0 & 0.4\\
        AP11 & 10.0 & 0.5\\
        AP12 & 11.0 & 0.5\\
        \hline
    \end{tabular}
    \caption{The aperture arrangements used in the experiment and analysis, with slit spacing and slit widths for each. These apertures were used in each of the five wavelengths of light, with the exception of $\lambda=1310$ nm, which does not use AP11 or AP12, and $\lambda=2000$ nm, which does not use AP1.}
    \label{tab:apertures}
\end{table} 

The experiment was performed in the Detector Characterization Laboratory at NASA Goddard Space Flight Center in January 2021. The test setup consists of a liquid nitrogen cryostat, a four-channel laser source, Sensor Chip Assembly (SCA) measurement and control electronics, temperature controllers, a controller for the aperture wheel stepper motor, and a controller for the cryogenic actuator to produce laser dither. The cryostat and an interior diagram are shown in Figure~\ref{fig:setup}.

\begin{figure*}[t]
\centering
\includegraphics[width=\linewidth]{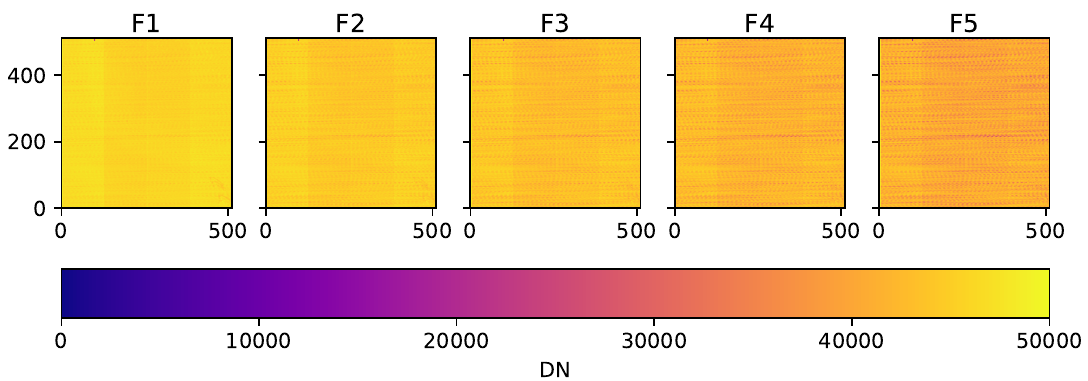}
    \caption{These panels show a $512\times512$ pixel portion of the entire array's display. Each panel is an individual frame of the speckle pattern realizations over time, F1 being the earliest and F5 the latest. The signal, measured in data numbers (DN), builds up between each frame and the interference fringes tend towards zero. The large-scale structure is consistent between each frame, with the patterns becoming more prominent in the later frames.}
    \label{fig:speckle}
\end{figure*}

\begin{figure*}[t]
    \centering
    \includegraphics[width=\linewidth]{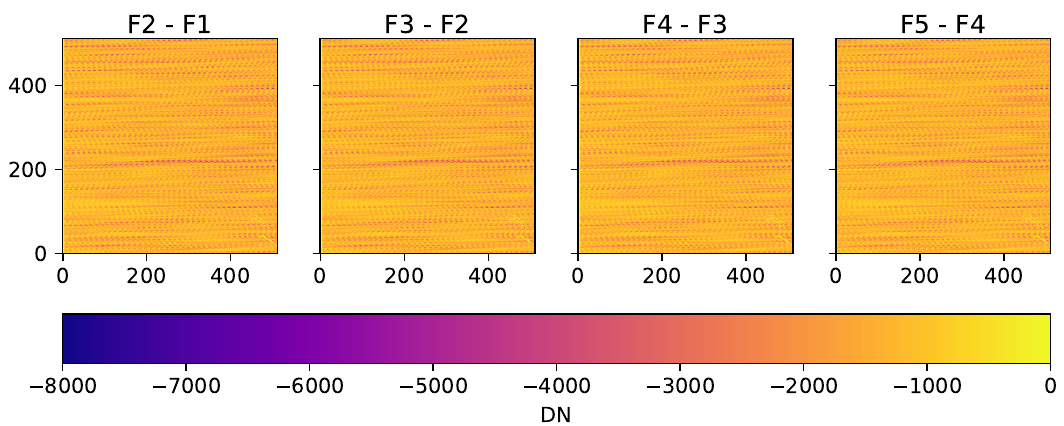}
    \caption{These panels show the same region of the detector as represented in Fig. \ref{fig:speckle}. The panels show the difference in the speckle pattern between each frame of measurement, displaying a consistent negative change between each frame due to the increasing nonlinearities in the detectors over time.}    
    \label{fig:speckle_differences}
\end{figure*}

Laser light is directed through the fiber into an integrating sphere to generate speckle patterns. The speckle patterns then illuminate a double-slit aperture, pass through a linear polarizer, and project interference patterns onto one chip of the detector array, rather than the full focal plane. The detector array used for this experiment was SCA 21536, which was qualified for mission use and included in the flight lot, but was not selected for flight \citep{Hirata2022}.

The data was collected using a sequence of twelve apertures with increasing slit separations across five wavelengths, $\lambda$: 850 nm, 980 nm, 1310 nm, 1550 nm, and 2000 nm. The aperture names, spacings, and widths used are reported in Table \ref{tab:apertures}. For every combination of aperture and wavelength, 25 speckle pattern realizations were generated and recorded.\footnote{We determined based on the fringe spacings that in the case of the 2000 nm data, the aperture number reported in the script is offset by one from the the aperture identified by the fringe spacings. We attribute this to the starting position of the aperture wheel in the 2000 nm run being offset by one position. }
Each speckle pattern realization includes five up-the-ramp measurements taken over a period of time, with 2.83 s between samples, allowing for a build up of data across the array \citep{Hirata2022}. This build up is measured across five frames of data, seen in Fig. \ref{fig:speckle}. The speckle patterns show a consistent change between each measurement, displayed in Fig. \ref{fig:speckle_differences}. This indicates the detector behaves mostly linearly, as the signal grows proportionally to the incoming light.

For later frames, we observe some amount of nonlinearity in the speckle pattern change between frames. Due to imperfections in the detectors, we expect some nonlinearity in the detector response over time, where the output signal is not proportional to the input. The presence of classical nonlinearities suggests that as signal builds up, there will be a reduction in contrast of fringes, which is seen in the data. Due to the evidence of compounding effects due to nonlinearities present in the detector, this analysis focuses majorly on the earlier frames, where the nonlinearities are small.

\section{MTF Measurement} \label{sec: MTF}
\subsection{Qualitative Theory of Speckle Autocorrelations}

The interference patterns of the speckle patterns projected onto the detector are dependent on the wavelength, $\lambda$, the distance between the aperture and the detector, $z_0$, and the spacing between the double slits, $d$. It is well accepted that the power spectrum of the speckle irradiance is proportional to the autocorrelation of the aperture \citep{Sensiper1993}. A full derivation of the relation between the observed speckle pattern, the detector modulation transfer function, and the slit parameters is given in Appendix~\ref{appendix_A}.

\begin{figure}[t]
    \centering
    \includegraphics[width=\linewidth]{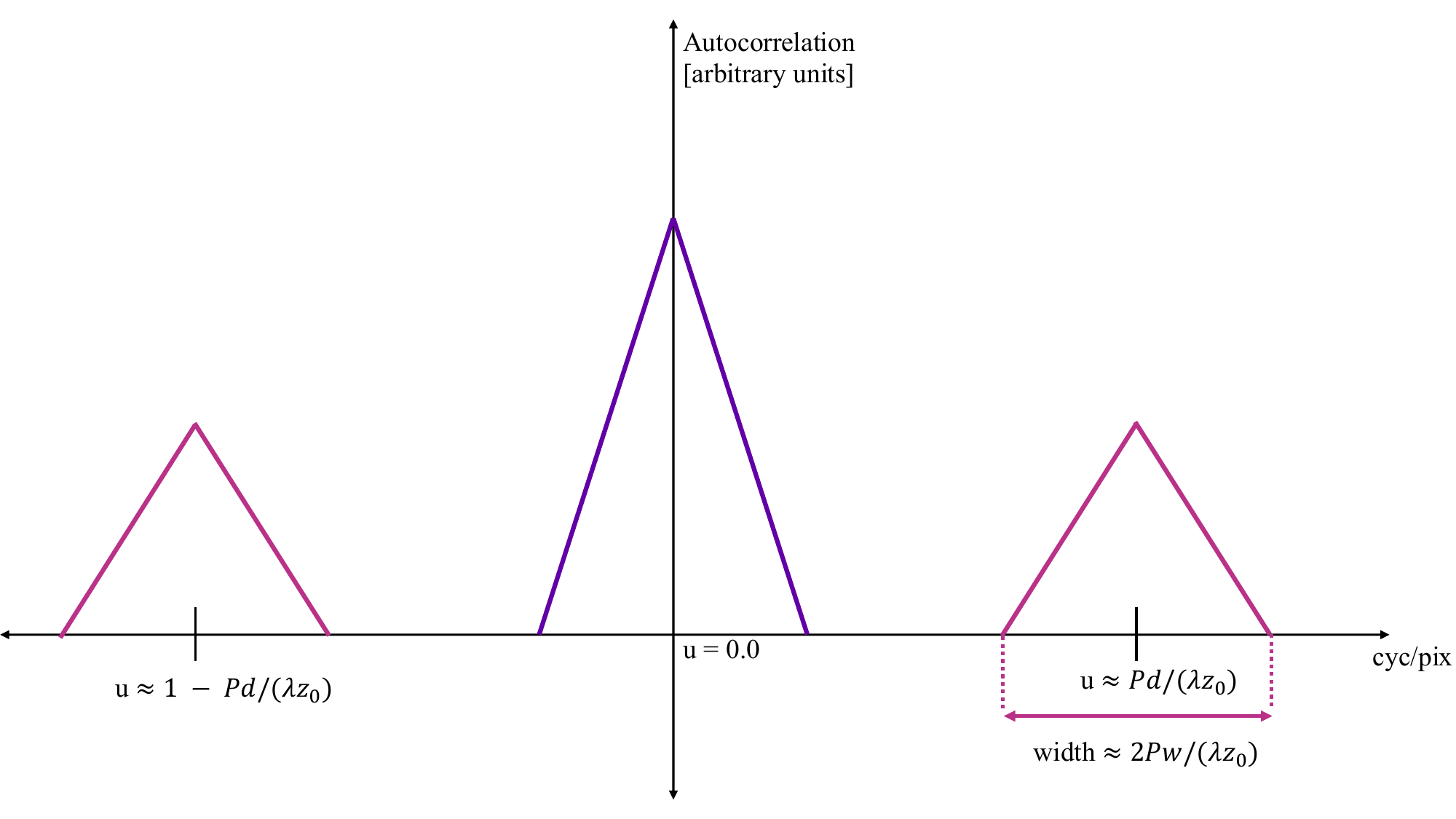}
    \caption{Visual representation of the autocorrelation of two top hat functions. The center triangle is centered at a spatial frequency of $u=0$ cyc/pix. The outer two triangles are the same width and about half the height of the center triangle. They are centered at the spatial frequencies of the fringes of the pattern. Due to the periodic boundaries in the Fourier domain, the spatial frequency of the left peak could be expressed either as $-Pd/(\lambda z_0)$ (the physical location) or $1-Pd/(\lambda z_0)$ (where it appears in a standard discrete Fourier transform).}
    \label{fig:triangles}
\end{figure}

\begin{figure*}[]
    \includegraphics[width=\textwidth]{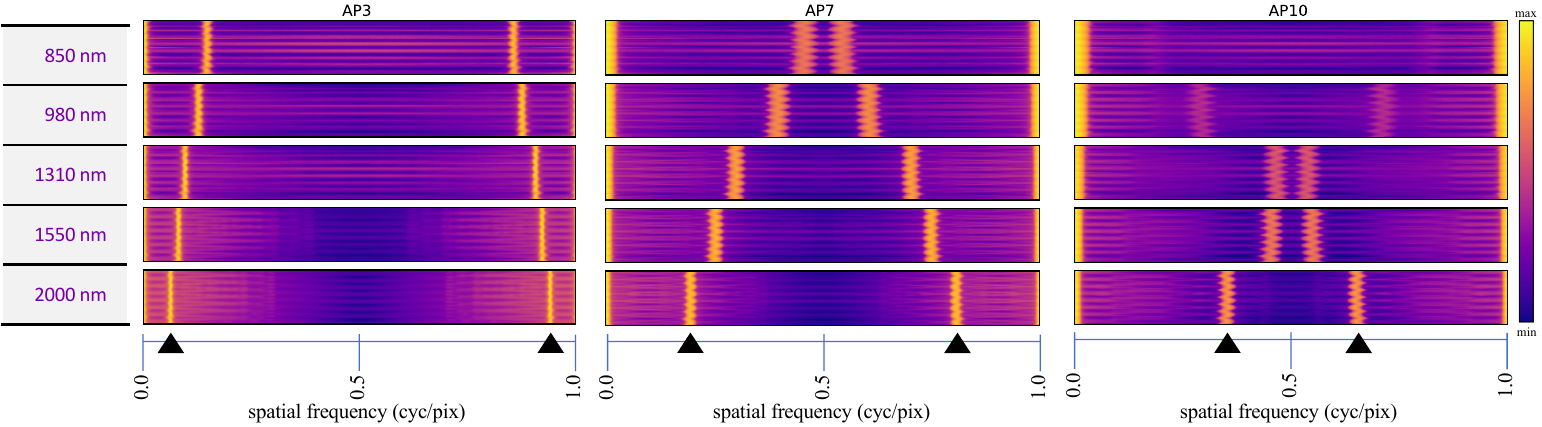}
    \caption{Examples of the 1D power spectra, in log scale, of the speckle data in three aperture arrangements: AP3, AP7, and AP10. For each aperture, we show 5 panels ranging from the bluest wavelength (850 nm, top) to the reddest (2000 nm, bottom). Within each panel, there are 64 rows of data indicating the $512\times 512$ subregions in the detector array. The plot ranges from $u=0$ to $u=1$ cyc/pix on the horizontal axis. The vertical bands at $u \approx Pd/(\lambda z_0)$ and $1-Pd/(\lambda z_0)$, marked with the black triangles, represent the interference fringes from the two slits. Each sub-panel is normalized separately since the flux per pixel varies depending on the laser source and slit width.}
    \label{fig:powerspectrum}
\end{figure*}

The double-slit aperture is a superposition of infinitesimal double slits, where each slit gives a true sine wave. It can be modeled as two top hat functions. The autocorrelation of two top hats is a set of triangles, one centered triangle and two outer triangles at approximately half the height of the center, illustrated in Fig. \ref{fig:triangles}. The center triangle is centered at spatial frequency $u = 0$, and the other two triangles are centered at the spatial frequencies of the fringes of the speckle pattern. The power spectrum of the autocorrelation, shown in Fig.~\ref{fig:powerspectrum}, will have corresponding peaks, each relating to the spatial frequencies of the fringes. (Note that the spatial frequency $u$ measured in a discrete Fourier transform has periodic boundary conditions, e.g., $u=-0.1$ wraps around to $u=0.9$.) A sequence of dots, as seen in Fig. \ref{fig:speckle_differences}, varies across the detector in distinct patterns. These patterns then directly translate to peaks in the power spectrum in Fourier space with widths that correlate to the length of the features. While there are significant features that fade within their own structures, there is also large scale variance across the region. This behavior will correspond to a peak present at a spatial frequency of zero. 

The MTF of a detector will suppress the power of some of the Fourier modes expected from the double slit experiment. The feature of the power spectrum centered at a spatial frequency of zero is unsuppressed by the MTF, while the other peaks are. We quantify the MTF by comparing the amplitude of the peaks of the power spectrum. The unsuppressed central peak is directly compared to the outer peaks, which occur at the fringes, to measure the MTF$^2$. From this value we return the MTF for each aperture arrangement. 

\subsection{Measurement} \label{sec:measurement}

The measurement of the MTF is performed by fitting the autocorrelation of the aperture to the power spectrum of the data. The autocorrelation measures how the features of the aperture will align at different offsets, describing the spatial frequency response of the system. We fit a model with a template of triangles, utilizing an estimation of the fringe locations, to the spatial frequencies and the height of the peaks from the power spectrum. We measured the amplitudes that correspond to the fit spatial frequency and compared them to obtain a value of the MTF$^2$. \textsc{solid-waffle} is a repository designed for the analysis and characterization of HxRG detectors at basic and advanced levels \citep{Hirata2019}, measuring features such as the brighter-fatter effect and nonlinearities. The \textsc{solid-waffle} characterization pipeline established the method to measure the MTF$^2$ and was originally tested with speckle data in the 1550 nm wavelength. This analysis makes use of these existing tools, performing the same measurements in more wavelengths and developing additional measurement methods as discussed further in this section. 

\begin{figure*}[t]
    \centering
    \includegraphics[width=6.5in]{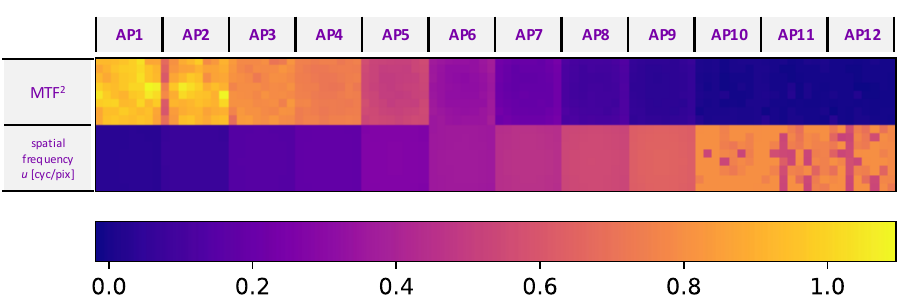}
    \caption{Visualization of the binning of the MTF$^2$ values and the spatial frequencies, u, using the original fitting method, measured in the 850 nm wavelength. The top row represents the MTF$^2$ values, the bottom row represents the fitted spatial frequencies, and each column represents a different aperture arrangement (AP1 through AP12). Each sub-panel shows results from an $8\times 8$ grid of regions on the detector array, each $512\times 512$ pixels.}
    \label{fig:datavis}
\end{figure*}

Using the established method, we produced a measurement of the MTF$^2$ across a $512\times512$ pixel region of the array. The array was split into 64 regions, producing an $8\times8$ grid for each aperture. In each region the MTF$^2$ was measured and we used the central value of the array for analysis. A visualization of the data in the 850 nm wavelength is presented in Fig. \ref{fig:datavis}. The majority of regions appear uniform with the exception of the MTF$^2$ in AP1 and 2, and the spatial frequency in AP10---12. The structure seen in AP1 and 2 is likely a result of fitting to very thin slits (hence thin triangles), which correspond to long-range correlations in real space (going out to $\lambda z_0/Pw \approx 200$ pixels). This makes the fit much more sensitive to unmodeled large-scale effects such as the flat field and channel-to-channel variations. The spatial frequency structure in AP10---12 is a result of the low MTF$^2$ and hence signal-to-noise ratio at high $u$; in an individual $512\times 512$ region, a free fit is easily confused by noise features. The features in the MTF$^2$ in AP1 and 2 exist in all wavelengths, but only 850 nm and 980 nm contain the nonuniform structure in spatial frequency in AP10---12, likely because they probe the largest spatial frequencies.

We measured the MTF$^2$ in the five wavelengths used in the experiment. For each, an estimation of the spatial frequency at which the fringes of the speckle pattern are expected was calculated using 
\begin{equation}
    u_{\text{predicted}} = \frac{Pd}{\lambda z_0}
\end{equation}
where $P$ is the pixel scale, $10\mu$m for \textit{Roman}, $d$ is the slit spacing between the aperture slits, $z_0$ is the distance between the aperture and the detector, and $\lambda$ is the wavelength of the light. 

Using the estimates of spatial frequency of the fringes, we fit the data to a model of the autocorrelation triangles, represented in Fig. \ref{fig:triangles} in each wavelength for every available aperture arrangement. At high spatial frequencies, the features in the power spectrum are faint, which prevents the model from accurately fitting the MTF. This can be seen in the power spectra of A10 in Fig. \ref{fig:powerspectrum}: the peaks of the triangles are difficult to discern visually from the background for the shorter wavelengths of 850 nm and 980 nm. To mitigate the effects of the low signal-to-noise ratio (SNR) in these wavelengths, we did a forced fit (see Appendix \ref{appendix_A}) to the predicted spatial frequencies in the model. This method avoids the issue that there is not enough SNR to find the correct peak in a $512 \times512$ pixel region, which is an issue at the largest $u$ due to the low signal-to-noise when the MTF is small. Also at $u \approx 0.5$, the peak at positive $u$ and the alias of the peak at $-u$ blend into a single feature centered at $u=0.5$, so a fit to a center and width does not work. Thus we perform the forced fit method at spatial frequencies above above $u=0.45$. To ensure the predicted values were accurate for the model, we utilized a distortion matrix of fringe spacing over the detector to account for non-small angle effects. The model in the forced fit also includes contributions from the Poisson noise of the photons and the background noise of the detector. The forced fit produced more accurate measurements of the MTF$^2$ in the regions of low SNR, resulting in more precise values for all of the experimental data. We performed this measurement of the MTF$^2$ for frames 1-2 (F1-2), 1-3 (F1-3), and 1-5 (F1-5) to understand how the contribution of charge diffusion changes with increasing effects from nonlinearities.  

\section{Charge Diffusion Model Fitting} \label{sec:modelfit}

In this section, we discuss the process of fitting the charge diffusion model to the measured MTF. First, the special cases are addressed, then the general, empirical model in Sec \ref{subsec:cd-models}, including a brief discussion on the process of fitting the model parameters. A discussion of the final results is in Sec \ref{subsec:model-results}. 

\subsection{Models}
\label{subsec:cd-models}

The MTF describes the pixel response function, or how a pixel responds to incoming light as a function of position on the detector. The ideal pixel response function is a top hat, discussed in Appendix \ref{appendix_A}, indicating that the pixel responds uniformly to light across its area and does not respond to external light. Because the detector is not perfect, we convolve the ideal PSF with other characteristics of the detector. The full MTF encompasses a combination of these additional contributions. For the purposes of this work, we consider the contributions of charge diffusion and inter-pixel capacitance (IPC), as both are likely to affect the measurement of the MTF. We model the MTF of the detector as

\begin{equation}
    \text{MTF}_{\text{system}} =  \text{MTF}_{\text{top\,hat}} \times \text{MTF}_{\text{charge \,diffusion}} \times \text{MTF}_{\text{IPC}}
\end{equation}

Charge diffusion's contribution to the MTF can be modeled in several ways. While the purpose of this paper is to establish a general, empirical model, we first consider two simple cases to illustrate physical limits of the model. Charge diffusion is dependent on the mobility and diffusion coefficients, which are unique characteristics of a detector. The mobility coefficient describes how quickly charges will move through the detector under the presence of an electric field. The diffusion coefficient describes how charges will diffuse over time due to random motion, without the presence of an electric field. We assume both of these to be constant throughout the detector. The general MTF model combines the mobility and diffusion terms with the top hat and IPC contributions to consider how drift and diffusion individually describe the charge diffusion. For the special cases, we consider when the mobility term and the diffusion term dominate the charge diffusion, respectively. 

In the drift-dominated case, the charge diffusion can be modeled using a Gaussian PSF \citep{Mosby2020}.
\begin{equation}
    \text{MTF}_{\text{drift}} = e^{-2\pi^2u^2 \sigma_G^2} 
\end{equation}
 When fit to the full range of MTF$^2$ data, the charge diffusion width is found to be $\sigma_G = 0.3140^{+0.0064}_{-0.0063}$ pix. The Gaussian model of the charge diffusion describes the MTF well at low spatial frequencies, but systematically underestimates the MTF at high spatial frequencies. Due to the exponential decay of the Gaussian tail, this model does not completely describe the behavior of charge diffusion in the detectors. 
 
In the diffusion-dominated case, the charge diffusion takes the form of a  hyperbolic secant function \citep{Holloway1986}. 
\begin{equation}
    \text{MTF}_{\text{diffusion}} = \text{sech}(2\pi u \sigma_S)
\end{equation}
When this model is fit to the data, the width of the function is found to be $\sigma_S = 0.3279^{+0.0043}_{-0.0042}$ pix. This model describes the data more accurately than the Gaussian at high spatial frequencies, and is a better estimated model for the MTF when the general model we derive in the next section is not used. However, there remains a range of frequencies for which this model is not a perfect predictor of the MTF.

These special cases do not fully describe the entire data set, motivating the development of a more accurate charge diffusion model. Assuming the diffusion and mobility terms to be constants, we derived a model of charge diffusion thoroughly in Appendix \ref{appendix_B}. With this model, we combine several physical parameters of the experiment into two unknown parameters: $h$, the absorption location of the holes, and $\xi = \frac{\mu E_z h}{D}$, which includes the grown-in electric field, $E_z$, the mobility term, $\mu$, and the diffusion coefficient, $D$, and is dimensionless. The general model of the charge diffusion contribution to the MTF is written in terms of $h$ and $\xi$: 

\begin{equation}
\begin{aligned}
    \text{MTF}_{cd} &= \int_{0}^{h}\frac{\alpha e^{-\alpha (h-z_0)}}{1-e^{-\alpha h}}\\
    &\Biggl[\frac{(e^{r_+z_0}-e^{r_-z_0})(r_-^2e^{r_-z_0}e^{h(r_+-r_-)} - r_+^2e^{r_+z_0})}{r_+e^{r_+z_0} - r_-e^{r_-z_0}e^{h(r_+-r_-)}} \\ 
    & + r_+e^{r_+z_0} - r_-e^{r_-z_0}\Biggr]^{-1}(r_+ - r_-)dz_0
    \label{MTF}
\end{aligned}
\end{equation}

where $\alpha$ is as described in \citep{Mosby2020}. The variables $r_+$ and $r_-$ are defined as:
\begin{equation}
    r_{\pm} = \frac{\xi}{2h} \pm \frac{\sqrt{\frac{\xi^2}{h^2}+16\pi^2(u^2+v^2)}}{2}
\end{equation}

By fitting this general model to the data, we determine how each parameter affects the image quality due to the intrinsic nature of the \textit{Roman} detectors.

\begin{figure*}[t]
    \centering
    \includegraphics[width=7in]{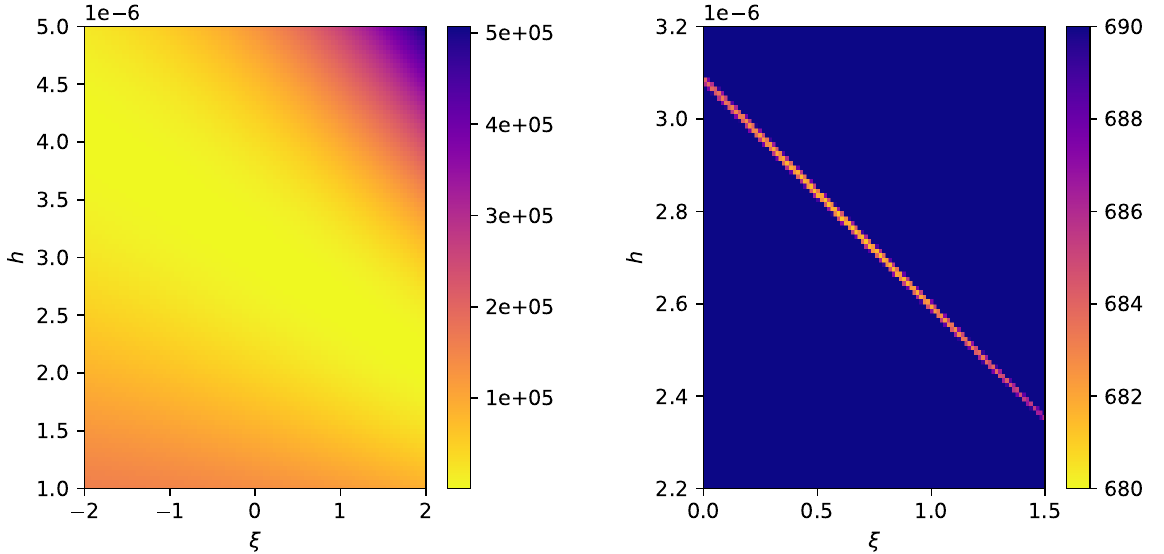}
    \caption{Contour plots of $\chi^2$ for a range of values for the parameters $h$ and $\xi$ in the general model for F1-2. The left plot showcases a complete view of the valley centered at a value of $\xi$ = 0. The right plot is a zoomed in and rescaled view, focused on the genuine minimum $\chi^2$ value. These plots were created using the data from the first two frames of the experiment.}
    \label{fig:f2_contour}
\end{figure*}

We first fit the $h$ and $\xi$ parameters using chi-squared minimization. Minimizing the model with respect to $h$ and $\xi$ proved to be difficult due to the shape of the $\chi^2$ valley seen in the left plot of Fig. \ref{fig:f2_contour}. Once the scale of the $\chi^2$ contour plot was stretched significantly, the real minimum was clear. This degenerate nature of the parameters necessitated a nested minimization method, motivated by the approximate minima we found by analyzing the chi-squared valley in Fig. \ref{fig:f2_contour} to find the best values for the model. We performed this model fitting for F1-2, F1-3, and F1-5. 

\subsection{Final Model Results}
\label{subsec:model-results}

The final results of the minimization are reported in Table \ref{tab:parameters}. In each frame, a positive $\xi$ value is reported. While this result is unphysical (see Appendix \ref{appendix_B}), $\xi = 0$ is within the 2$\sigma$ range for F1-2. This result indicates the grown-in electric field, which is unique to the H4RG-10 detectors, is not significantly detected in the early frames where nonlinearities are insignificant. Thus, we believe the value of $\xi$ is not significantly positive and is not concerning. As the frames increase, $\xi$ trends positive. This is likely a result of increasing nonlinearities in the detector over time. There is no complete model of nonlinear effects, such as the brighter-fatter effect, at the time of this work. This will be explored in the future.

\begin{figure*}
    \centering
    \includegraphics[width=6in]{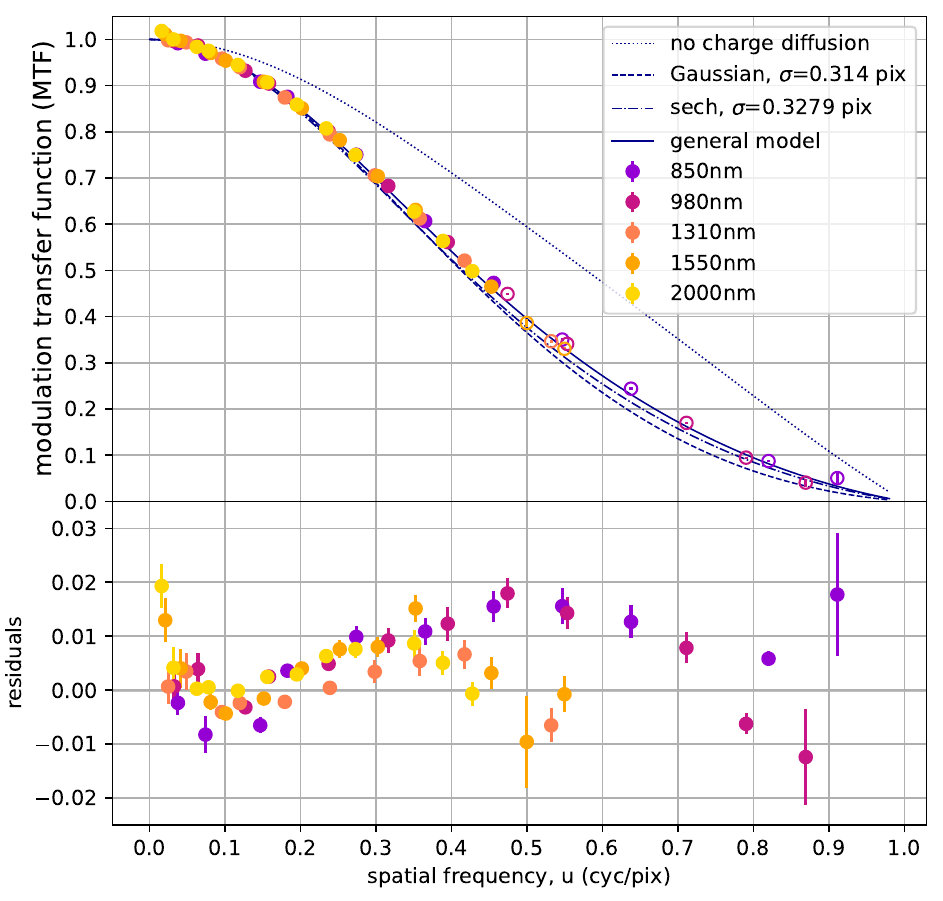}
    \caption{\label{fig:F2-full} MTF measurements acquired from SCA 21536 using the data from the first two frames of the experiment. Each point represents the mean value across the detector array. Solid circles show the fit of the MTF and spatial frequency, while open circles indicate the points where a forced fit (see Appendix \ref{appendix_A}) was used. The general charge diffusion model (this work) is shown as a solid line. Special case models and the model of the MTF without a contribution from charge diffusion are shown as various dotted lines. The residuals show periodic behavior with a discrepancy around $u = 0.5$ cyc/pix.}
\end{figure*}

Using the values presented in Table \ref{tab:parameters}, our general, empirical model is plotted along with the measured data for F1-2 in Fig. \ref{fig:F2-full}. We also plot the MTF model without a charge diffusion contribution, and the special cases for reference. The figure shows that the general model we derived in this work fits the data well at the entire range of spatial frequencies, an improvement from the special case fits. 

Figure \ref{fig:F2-full} also shows residuals from our general MTF model. The residuals display periodic behavior in all frames, indicating that our MTF model is missing some important contributions. We discuss this result further in Sec. \ref{sec:discussion}. 

\begin{table}[h]
    \begin{tabular}{|c|c c|}
        \hline
        Frames & $\xi$ & h \\ & (dimensionless) & ($\mu$m) \\
        \hline
        F1-2 & 0.64$_{-0.38, 0.77}^{+0.39, 0.79}$ & 2.77$_{-0.19, 0.39}^{+0.19, 0.38}$  \\ 
        F1-3 & 1.78$_{-0.40, 0.78}^{+0.43, 0.91}$ & 2.28$_{-0.21, 0.43}^{+0.20, 0.39}$\\
        F1-5 & 2.68$_{-0.47, 0.90}^{+0.51, 1.1}$ & 1.98$_{-0.24, 0.48}^{+0.23, 0.45}$\\
        \hline
    \end{tabular}
    \caption{Best fit parameters of $h$ and $\xi$ for each frame. The result of $\xi$ for F1-2 is consistent with zero within 2$\sigma$, meaning the effects of the grown-in electric field is not significantly detected in the early frames where nonlinearities are insignificant. The F1-3 and F1-5 results indicate a significant presence of nonlinearities resulting in the more extreme values of $\xi$.}
    \label{tab:parameters}
\end{table} 

Additionally, at $u=0.5$ cyc/pix, there is a distinct separation between the wavelengths in the residuals. This feature is present in the residuals for all of the modeled frames, the special case residuals, and the MTF without charge diffusion. Since it is a persistent feature, this can likely be attributed to another effect in the detector that impacts the MTF besides the charge diffusion. 

\section{Discussion} \label{sec:discussion}

Our general, empirical model of the charge diffusion as a contribution to the MTF fits the data well, as indicated by the small residuals. In the early frames of the data, we do not detect a significant value for $\xi$ since it is consistent with zero. This indicates that the grown-in drift field, $E_z$, in the HgCdTe detectors is not causing a measurable deviation from the sech profile. 

\begin{figure}
    \centering
    \includegraphics[width=\linewidth]{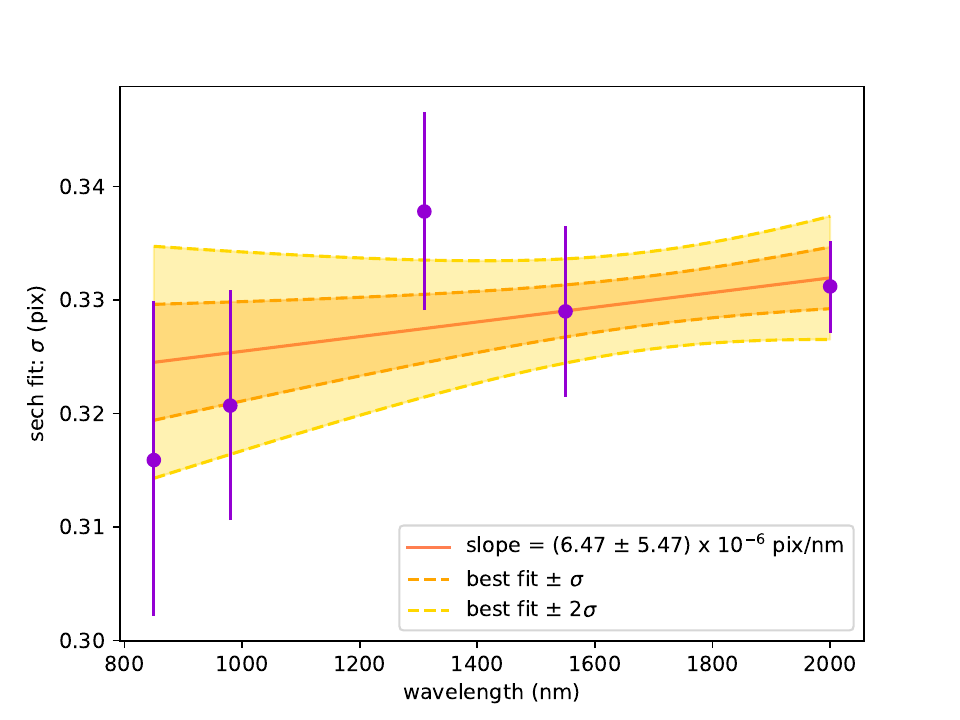}
    \caption{Dependence of the width of the diffusion-dominated case of charge diffusion on the wavelength. All five experimental wavelengths are used, and a line was fit to the data. The shaded regions display the error of the fit, with the light outer region within 2$\sigma$ and the dark inner within 1$\sigma$. The slope of the line is consistent with zero within 2$\sigma$, indicating a lack of dependence on wavelength.}
    \label{fig:wavelengthdependence}
\end{figure}

Despite the lack of wavelength contribution in the model, we wanted to confirm the wavelength independent nature of charge diffusion. Since there is not a significant contribution from $\xi$, we tested the wavelength dependence using the diffusion-dominated special case. The width of the model, $\sigma$, was fit for each wavelength of data independently. These values are plotted against the wavelengths and we fit a line to the data, shown in Fig. \ref{fig:wavelengthdependence}. The slope of this line is consistent with zero within 2$\sigma$, confirming the lack of wavelength dependence within the MTF model. This indicates that the same charge diffusion model can be applied in all wavelength measurements. 

The residuals from the general MTF model exhibit periodic behavior, indicating an incomplete model. This behavior is present in all frames, and in special cases of the MTF model in Fig. \ref{fig:F2-full}. Due to the persistence of the periodic residuals, it is likely due to other non-linear effects not modeled in this project. Future works will seek to characterize these potential sources of non-linear effects, including the brighter-fatter effect and a more developed version of the inter-pixel capacitance. Despite this, the residuals are fairly small and consistently centered around zero, indicating this model for charge diffusion can be confidently used in other efforts regarding the \textit{Roman} detectors. 

Our final recommended model is therefore the sech model from the F1-2 fits, which have the smallest non-linearity contribution. This has $\sigma = 0.3279^{+0.0043}_{-0.0042}$ pix. We estimate that the dominant systematic error is from remaining non-linearities; the F1-3 fits, which go to twice the fluence, have a $\sigma$ that is larger by 0.0093 pix (0.3372 vs.\ 0.3279), and we recommend this as the systematic uncertainty.

A previous study of the charge diffusion in the \textit{Roman} detectors determined the charge diffusion using pairwise pixel probability for the Gaussian special case \citep{Givans2022}. The measurement of charge diffusion was completed for three SCAs different from the one used in this project: 20663, 20828, and 20829. The reported charge diffusion lengths were 0.27, 0.35, and 0.33 pix respectively, for a wavelength of 500 nm. These results are comparable to the charge diffusion lengths of $\sigma_S$ and $\sigma_G$ determined through this project for the Gaussian and hyperbolic secant special cases, specifically the fits that include more nonlinear effects. As this project only included measurements in one wavelength, further investigation into the wavelength dependence of the charge diffusion was necessary.

The measurement of charge diffusion in this project has important implications for the strategy of the \textit{Roman} High Latitude Wide Area Survey (HLWAS). Since charge diffusion occurs before pixelization, a PSF with charge diffusion is better sampled than one without \citep{Mosby2020}. In the case of \textit{Roman}, which is diffraction-limited in the Y106 band filter, the reported charge diffusion will effectively smear the image to the resolution of the J129 filter \citep{Hirata2024}. With the confirmed presence of significant charge diffusion, the Y106 filter will be sufficiently sampled and thus was recommended for the survey's medium and deep tiers. 

The charge diffusion model for \textit{Roman} is crucial for the development of simulations, software development, and general planning of the mission. A physical understanding of how photons convert to output counts is necessary for simulations used to test the image processing tools being prepared; charge diffusion is one element of this process due to its contribution to the PSF. Charge diffusion was modeled using the hyperbolic secant model with $\sigma = 0.3279$ pix for the simulated sky produced in OpenUniverse2024 \citep{Troxel2025}. The new model of charge diffusion derived in this work is being applied to current efforts in cosmic ray and PSF modeling. In the cosmic ray simulations, the charge diffusion term of the MTF will be used to support the mapping of spatial distributions of electrons in the HgCdTe material (Harbo Torres et al., in prep). A suite of simulations for testing and modeling the PSF fitting code will also include the charge diffusion model.

Through the determination of the charge diffusion's contribution to the MTF, we improve characterization of the \textit{Roman} detectors. This allows for increased accuracy in models of the PSF and thus improved precision for weak lensing measurements. As our ability to measure weak lensing improves, the large-scale structure of the universe will be mapped to greater extent with greater amounts of detail. This will provide insight into the growth of cosmic structure and the characterization of dark energy and the accelerated expansion of the universe. 

\section*{Acknowledgments}

We thank Ami Choi, Nihar Dalal, and Anthony Harbo Torres for providing their continued support and feedback throughout the duration of this project.

This paper is based on data acquired at the Detector Characterization Laboratory at NASA Goddard Space Flight
Center.

This work made extensive use of the Pitzer and Cardinal clusters at the Ohio Supercomputer Center.

This work was supported by the NASA ROSES grant 22-ROMAN11-0011, contract number 80NM0024F0012, via a JPL subaward.

\appendix
\section{Theory of Speckle Autocorrelations}
\label{appendix_A}

This appendix presents a re-derivation of the speckle method from first principles. Although the speckle autocorrelation method is well-established in the literature \citep{Sensiper1993, Pozo2005}, questions arose during this project on (i) the large-angle distortions, and (ii) when sinc versus sinc$^2$ functions should be used to describe the pixel tophat. The full derivation was helpful in answering these questions within the {\slshape Roman} project, and we hope it will be useful to future users of the speckle technique. The formalism is consistent with that used in \citet{Hirata2022}.

\subsection{Speckle patterns}

Let's suppose that we have a speckle pattern generated from an aperture ${\cal A}$ at some distance $z_0$ from the detector. We will put the center of the aperture (for purposes of Taylor expansions) at $(0,0,z_0)$. Let's further suppose that we are working with only one polarization of light (so we will suppress polarization indices), and it has a wavelength $\lambda = 2\pi/k$.

\begin{figure}[h]
\centering{\includegraphics[width=3.75in]{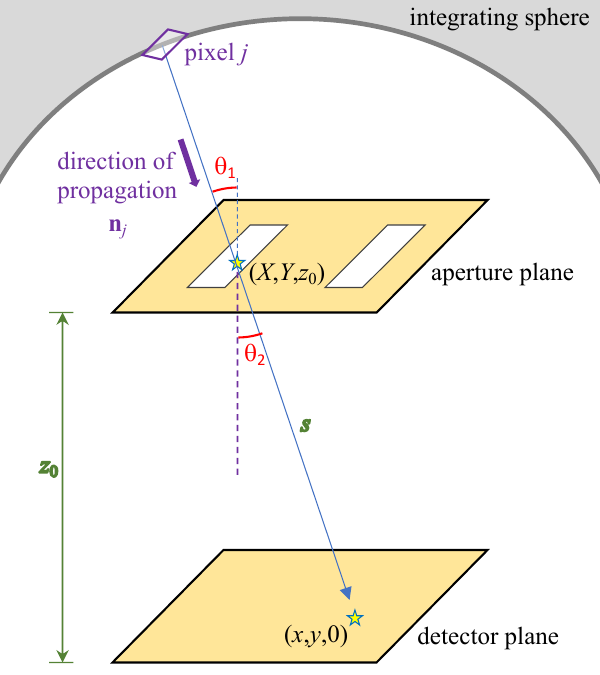}}
\caption{\label{fig:specklefig}The geometry and symbols used in Appendix~\ref{appendix_A} (not to scale). The diffusing inner surface of the integrating sphere is divided into pixels $j$. Light from each pixel passes through the aperture plane, and then on to the detector plane.}
\end{figure}

The light arriving at the aperture can be treated as a superposition of plane waves. If there is a single incident plane wave $E({\bf r}) = E_0 e^{ik\hat{\bf n}\cdot{\bf r}}$ (where $\hat{\bf n}$ is the direction of propagation and $E_0$ is the complex amplitude), then the Fresnel-Kirchhoff diffraction formula (see, e.g., \citealt{1999poet.book.....B}, 8.3.17) gives
\begin{equation}
E(x,y,0) = -\frac{ikE_0}{4\pi} \int_{\cal A} \frac{e^{ik\hat{\bf n}\cdot(X,Y,z_0)} e^{iks}}{s} (\cos\theta_1 + \cos\theta_2)\,dX\,dY,
\label{eq:FK}
\end{equation}
where the capital $X$ and $Y$ denote lateral coordinates in the aperture; $\cos\theta_1 = -n_z$ is the angle of incidence of the plane wave on the aperture, $\cos\theta_2 = z_0/s$ is the angle of exit, and the distance $s$ is
\begin{equation}
s = \sqrt{(x-X)^2 + (y-Y)^2 + z_0^2}.
\end{equation}
(See Fig.~\ref{fig:specklefig} for the geometry.)
To generate a speckle pattern, we illuminate the aperture with uniform specific intensity from above, i.e., a superposition of many plane waves each with a direction of propagation $\hat{\bf n}_j$, coming from a pixel $j$ on the integrating sphere that subtends a solid angle $\Omega_j$. For each speckle pattern realization, the electric field $E_{0j}$ of light emerging from that pixel in the integrating sphere is drawn from a complex Gaussian $\langle E_{0j} E_{0j'}^\ast\rangle_{\rm sp} = 2I_{\rm inc}\delta_{jj'} \Delta\Omega/(\epsilon_0c)$, where $I_{\rm inc}$ is the radiance. (A Gaussian is used in accordance with the central limit theorem.) The average here is an average over speckle realizations (within a single realization, the electric fields coming from different directions have a definite relative phase since they all come from the laser). We may obtain the electric field correlation function from Eq.~(\ref{eq:FK}), summing over all of these directions:
\begin{eqnarray}
\langle E(x,y,0) E^\ast(x',y',0) \rangle_{\rm sp}
&=& \sum_j \frac{2I_{\rm inc} \Delta\Omega}{\epsilon_0c} \left|-\frac{ik}{4\pi} \right|^2 
\int_{\cal A} \frac{e^{ik\hat{\bf n}\cdot(X,Y,z_0)} e^{iks}}{s} (\cos\theta_1 + \cos\theta_2)\,dX\,dY
\nonumber \\ && \times
\int_{\cal A} \frac{e^{-ik\hat{\bf n}\cdot(X',Y',z_0)} e^{-iks'}}{s'} (\cos\theta_1 + \cos\theta'_2)\,dX'\,dY'.
\label{eq:E-sum}
\end{eqnarray}
Here we have primed $s$ and $\theta_2$ in the second integral because they depend on the detector plane position $(x',y')$ and the aperture plane position $(X',Y')$.

We now replace the summation over solid angles with an integral over the range ${\mathbb S}$ of solid angles of illumination of the aperture:
\begin{eqnarray}
\langle E(x,y,0) E^\ast(x',y',0) \rangle_{\rm sp}
&=& \frac{k^2I_{\rm inc}}{8\pi^2\epsilon_0c} \int_{\mathbb S} d\Omega \,
\int_{\cal A}\int_{\cal A} \frac{e^{ik\hat{\bf n}\cdot(X-X',Y-Y')} e^{ik(s-s')}}{ss'}
\nonumber \\ && \times
 (\cos\theta_1 + \cos\theta_2)(\cos\theta_1 + \cos\theta'_2)\,dX\,dY\,dX'\,dY'.
 \label{eq:E-sum-temp}
\end{eqnarray}
We can see from the solid angle integral that the integral will be dominated by the contribution where $(X-X',Y-Y')$ is at most a few wavelengths (if the separation is greater than that, the solid angle integral is an integral of a rapidly oscillating function and averages to zero). We further make the approximation that $(x,y)$ and $(x',y')$ are close to each other. So we substitute the first-order Taylor expansion of $s-s'$ in $(x',y')$ and $(X',Y')$ around $(x,y)$ and $(X,Y)$:
\begin{equation}
s-s' \approx \frac{(x-X)(x-x'-X+X') + (y-Y)(y-y'-Y+Y')}s.
\end{equation}
We also keep the difference between $s$ and $s'$ only in the phase, and drop the small difference between $\cos\theta_2$ and $\cos\theta'_2$:
\begin{eqnarray}
\langle E(x,y,0) E^\ast(x',y',0) \rangle_{\rm sp}
&=& \frac{k^2I_{\rm inc}}{8\pi^2\epsilon_0c} \int_{\mathbb S} \,d\Omega \,
\int_{\cal A}\int_{\cal A} \frac{e^{ik\hat{\bf n}\cdot(X-X',Y-Y')} e^{ik[(x-X)(x-x'-X+X') + (y-Y)(y-y'-Y+Y')]/s}}{s^2}
\nonumber \\ && \times
 (\cos\theta_1 + \cos\theta_2)^2 \,dX\,dY\,dX'\,dY'.
\end{eqnarray}
The integral over $(X',Y')$ over a region near $(X,Y)$ is then an integral of a complex exponential, which is a $\delta$-function (assuming the aperture is many wavelengths wide so that we may drop edge effects):
\begin{eqnarray}
\langle E(x,y,0) E^\ast(x',y',0) \rangle_{\rm sp}
&=& \frac{k^2I_{\rm inc}}{8\pi^2\epsilon_0c} \int_{\mathbb S} \,d\Omega \,
\int_{\cal A} \frac{ e^{ik[(x-X)(x-x') + (y-Y)(y-y')]/s}}{s^2} (2\pi)^2 \delta\left( k\hat n_x - \frac{x-X}s \right)
\delta\left( k\hat n_y - \frac{y-Y}s \right)
\nonumber \\ && \times
 (\cos\theta_1 + \cos\theta_2)^2 \,dX\,dY.
\end{eqnarray}
The $\delta$ function sets $\cos\theta_1=\cos\theta_2$. Since the orthographic projection of the solid angle element is $d\hat n_x \,d\hat n_y = \cos\theta_1\,d\Omega$, the solid angle integral collapses and we have
\begin{equation}
\langle E(x,y,0) E^\ast(x',y',0) \rangle_{\rm sp}= \frac{2I_{\rm inc}}{\epsilon_0c} 
\int_{\cal A} \frac{ e^{ik[(x-X)(x-x') + (y-Y)(y-y')]/s}}{s^2}
\cos\theta_2 \,dX\,dY.
\label{eq:EE}
\end{equation}

What is usually of interest to us is the intensity and intensity fluctuations at the detector. Let us choose a small region of the detector and suppose that the aperture is small enough to neglect the variation of $\cos\theta_2$ and $s$ over the aperture. Then the realization average of the irradiance $F$ at the detector is
\begin{equation}
\bar F \equiv \langle F(x,y) \rangle_{\rm sp} = \frac12\epsilon_0c \cos\theta_2 \langle |E(x,y,0)|^2 \rangle_{\rm sp} = 
\frac{I_{\rm inc}A\cos^2\theta_2}{s^2},
\end{equation}
where $A$ is the area of the aperture. This is just the usual geometrical formula.

For speckle studies, we want the correlation function of the irradiance. This is a fourth moment of the electric field, and can be obtained from Wick's theorem:
\begin{eqnarray}
{\rm Cov}_{\rm sp} [F(x,y), F(x+\Delta x, y+\Delta y) ]
\!\!\!\!\!\!\!\!  \!\!\!\!\!\!\!\! \!\!\!\!\!\!\!\! \!\!\!\!\!\!\!\! \!\!\!\!\!\!\!\! \!\!\!\!\!\!\!\! && 
\nonumber \\
&=& \langle \langle F(x,y)F(x+\Delta x,y+\Delta y) \rangle_{\rm sp} - \langle F(x,y) \rangle_{\rm sp}\langle F(x+\Delta x,y+\Delta y) \rangle_{\rm sp}
\nonumber \\
&=& \left( \frac12\epsilon_0c \cos\theta_2 \right)^2 \bigl[ \langle E(x,y,0) E^\ast(x,y,0)E(x+\Delta x,y+\Delta y,0) E^\ast(x+\Delta x,y+\Delta y, 0) \rangle_{\rm sp}
\nonumber \\ && ~~~~
 - \langle E(x,y,0) E^\ast(x,y,0) \rangle_{\rm sp}\langle E(x+\Delta x,y+\Delta y,0) E^\ast(x+\Delta x,y+\Delta y, 0) \rangle_{\rm sp} \bigr]
\nonumber \\
&=&  \left( \frac12\epsilon_0c \cos\theta_2 \right)^2 \left[ \left| \langle E(x,y,0)E^\ast(x+\Delta x,y+\Delta y, 0) \rangle_{\rm sp} \right|^2
+ \left| \langle E(x,y,0)E(x+\Delta x,y+\Delta y, 0) \rangle_{\rm sp} \right|^2
\right]
\nonumber \\
&=& \frac{I_{\rm inc}^2 \cos^4\theta_2}{s^4} \left| \int_{\cal A} e^{ik[(x-X)\Delta x + (y-Y)\Delta y]/s} \,dX\,dY \right|^2
\nonumber \\
&=& \bar F^2 \left| \frac1A \int_{\cal A} e^{ik[(x-X)\Delta x + (y-Y)\Delta y]/s} \,dX\,dY \right|^2.
\label{eq:Cov-Fxy.temp}
\end{eqnarray}
(The correlation function with 2 electric fields and no complex conjugate is zero because when we average over realizations the phase is random.)
We thus see that the speckle correlation function is related to the square of the Fourier transform of the aperture. If the region of interest on the detector is not near $(0,0)$, then there is a subtlety in that $s$ also depends on $X$ and $Y$. To take this into account, we perform the Taylor expansion of $s$ to first order in $X$ and $Y$:
\begin{equation}
\frac1s = \frac1{\sqrt{x^2+y^2+z_0^2}} + \frac{xX+yY}{(x^2+y^2+z_0^2)^{3/2}} + ...
\,,
\end{equation}
where $\cos\theta_2$ is evaluated at $X=Y=0$. Then we can Taylor expand the exponent in $(X,Y)$, and Eq.~(\ref{eq:Cov-Fxy.temp}) becomes
\begin{equation}
{\rm Cov}_{\rm sp} [F(x,y), F(x+\Delta x, y+\Delta y) ] = \bar F^2 \left| \frac1A \int_{\cal A} e^{-ik (\Delta x,\Delta y)\cdot{\bf M}(X,Y)/z_0} \,dX\,dY \right|^2,
\label{eq:Cov-Fxy}
\end{equation}
where ${\bf M}$ is the $2\times 2$ distortion matrix
\begin{equation}
{\bf M} =
\left. \frac{\partial(-(x-X)/s, -(y-Y)/s)}{\partial(X,Y)} \right|_{(X,Y)=(0,0)}
= \frac{z_0}{(z_0^2+x^2+y^2)^{3/2}} \left( \begin{array}{cc}
z_0^2 + y^2 &  -xy \\  -xy & z_0^2+x^2
\end{array} \right).
\label{eq:matrix-M}
\end{equation}
This matrix is symmetric and is simply the $2\times 2$ identity if $(x,y)=(0,0)$, i.e., the region of interest is directly below the aperture. The matrix ${\bf M}$ is responsible for, e.g., the fringe spacing varying over the detector. Its determinant is
\begin{equation}
\det{\bf M} = \frac{z_0^4}{(z_0^2+x^2+y^2)^2} = \cos^4\theta_2.
\label{eq:detM}
\end{equation}

\subsection{Power spectral density of speckles on detector}

A detector has a grid of pixels centered at $(x,y) = (m_xP,m_yP)$, where $m_x$ and $m_y$ are integers, and $P$ is the pixel pitch (10 $\mu$m for {\slshape Roman}). We suppose that the pixels have a response function $R$, such that the signal $S$ in pixel $(m_x,m_y)$ is given by
\begin{equation}
S_{m_x,m_y} = \frac{\eta P^2 \Delta t}{g(hc/\lambda)} \int_{{\mathbb R}^2} R(\xi_x, \xi_y) F(m_xP+\xi_x, m_yP+\xi_y) d\xi_x \,d\xi_y,
\label{eq:Sxy}
\end{equation}
where $\eta$ is the quantum efficiency; $\Delta t$ is the exposure time; $g$ is the gain; $hc/\lambda$ is the energy per photon; and ${\boldsymbol\xi}$ is a 2D vector indicating how far from the pixel center a photon lands. The response function is normalized by
\begin{equation}
\int_{{\mathbb R}^2} R(\xi_x, \xi_y) \,d\xi_x\,d\xi_y = 1.
\end{equation}
An ``ideal'' detector would have a top-hat function:
\begin{equation}
R_{\rm ideal}(\xi_x,\xi_y) = \left\{\begin{array}{lll} 1/P^2 & & |\xi_x|<\frac12P ~~{\rm and}~~ |\xi_y|<\frac12P
\\ 0 & & {\rm otherwise}\end{array}\right.~~
\rightarrow~~
\tilde R_{\rm ideal}(u,v) = {\rm sinc}\left(Pu\right){\rm sinc}\left(Pv\right),
\end{equation}
where the $\tilde{~}$ denotes a Fourier transform; we have chosen to write ${\boldsymbol\xi}$ in units of physical length (m) and $(u,v)$ in units of inverse length (m$^{-1}$). The mean signal is $\bar S = \eta P^2\Delta t\,\bar F/(ghc/\lambda)$. The correlation function of the signal is
\begin{eqnarray}
{\rm Cov}_{\rm sp} [ S_{m_x,m_y}, S_{m_x+\Delta m_x, m_y+\Delta m_y} ]
&=& 2\bar S^2  \int_{{\mathbb R}^2}\int_{{\mathbb R}^2} d\xi_x \,d\xi_y \,d\xi'_x \,d\xi'_y\, R(\xi_x, \xi_y) R(\xi'_x, \xi'_y)
\nonumber \\ && \times
\left| \frac1A \int_{\cal A} 
e^{-ik (\xi'_x-\xi_x + P\Delta m_x,\xi'_y-\xi_y + P\Delta m_y)\cdot{\bf M}(X,Y)/z_0}
 \,dX\,dY \right|^2.
\end{eqnarray}
We may expand the last squared integral into an integral over $dX\,dY$ and an identical integral over $dX'\,dY'$ with a complex conjugate. Then we write $X'=X+\Delta X$ and $Y'=Y+\Delta Y$ to get
\begin{eqnarray}
\!\!\!\!\!\!\!\!
{\rm Cov}_{\rm sp} [ S_{m_x,m_y}, S_{m_x+\Delta m_x, m_y+\Delta m_y} ]
&=& \bar S^2  \int_{{\mathbb R}^2}\int_{{\mathbb R}^2} d\xi_x \,d\xi_y \,d\xi'_x \,d\xi'_y\, R(\xi_x, \xi_y) R(\xi'_x, \xi'_y)
\nonumber \\ && \times
 \frac1{A^2} 
 \int_{\cal A} \int_{\cal A} e^{ik (\xi'_x-\xi_x + P\Delta m_x,\xi'_y-\xi_y + P\Delta m_y)\cdot{\bf M}(\Delta X,\Delta Y)/z_0} \,dX\,dY  \,dX'\,dY'.
\end{eqnarray}
We define the normalized aperture autocorrelation function $T(\Delta X,\Delta Y)$ to be
\begin{equation}
T(\Delta X,\Delta Y) =  \frac1{A} 
 \int_{{(X,Y)\in \cal A}~{\rm and}~{(X+\Delta X,Y+\Delta Y)\in \cal A}} dX\,dY;
\end{equation}
this is an even function by construction, $T(\Delta X,\Delta Y) = T(-\Delta X,-\Delta Y)$, and has $T(0,0)=1$. Then
\begin{eqnarray}
{\rm Cov}_{\rm sp} [ S_{m_x,m_y}, S_{m_x+\Delta m_x, m_y+\Delta m_y} ]
&=& \frac{\bar S^2}{A}  \int_{{\mathbb R}^2}\int_{{\mathbb R}^2} d\xi_x \,d\xi_y \,d\xi'_x \,d\xi'_y\, R(\xi_x, \xi_y) R(\xi'_x, \xi'_y)
 \int_{{\mathbb R}^2} d\Delta X\,d\Delta Y\, T(\Delta X,\Delta Y) 
 \nonumber \\ && \times
e^{ik (\xi'_x-\xi_x + P\Delta m_x,\xi'_y-\xi_y + P\Delta m_y)\cdot{\bf M}(\Delta X,\Delta Y)/z_0} .
\end{eqnarray}
The integrals over $d\xi_x ... d\xi'_y$ can be pulled to the inside of the $d\Delta X\,d\Delta Y$ integral, and factored:
\begin{eqnarray}
{\rm Cov}_{\rm sp} [ S_{m_x,m_y}, S_{m_x+\Delta m_x, m_y+\Delta m_y} ]
&=& \frac{\bar S^2}{A}  
 \int_{{\mathbb R}^2} d\Delta X\,d\Delta Y\, T(\Delta X,\Delta Y) 
 \int_{{\mathbb R}^2}  d\xi_x \,d\xi_y\, R(\xi_x, \xi_y)
e^{ik (-\xi_x ,-\xi_y )\cdot{\bf M}(\Delta X,\Delta Y)/z_0} 
 \nonumber \\ && \times
  \int_{{\mathbb R}^2} \,d\xi'_x \,d\xi'_y\,  R(\xi'_x, \xi'_y)
e^{ik (\xi'_x,\xi'_y)\cdot{\bf M}(\Delta X,\Delta Y)/z_0}
 \nonumber \\ && \times
e^{ik (P\Delta m_x,P\Delta m_y)\cdot{\bf M}(\Delta X,\Delta Y)/z_0} .
\end{eqnarray}
The ${\boldsymbol\xi}$ and ${\boldsymbol\xi}'$ integrals are two copies of the Fourier transform $\tilde R$, one with a complex conjugate. Writing this in terms of $\lambda$ instead of $k=2\pi/\lambda$ then gives:
\begin{eqnarray}
{\rm Cov}_{\rm sp} [ S_{m_x,m_y}, S_{m_x+\Delta m_x, m_y+\Delta m_y} ]
&=& \frac{\bar S^2}{A} \int_{{\mathbb R}^2} d\Delta X\,d\Delta Y\, T(\Delta X,\Delta Y) 
\left| \tilde R\left( \frac{{\bf M}(\Delta X,\Delta Y)}{\lambda z_0} \right) \right|^2
\nonumber \\ && \times
e^{2\pi i P (\Delta m_x ,\Delta m_y)\cdot{\bf M}(\Delta X,\Delta Y)]/(\lambda z_0)} .
\label{eq:corr-sp}
\end{eqnarray}
Finally, we want to compute the PSD. We normally do this on an $N_x\times N_y$ region of pixels. The method is to subtract the mean (which will only affect the zero Fourier mode), take the discrete Fourier transform and then square it. The result is:
\begin{eqnarray}
\langle |\tilde S_{j_x,j_y}|^2 \rangle_{\rm sp} &=&
 \frac1{(N_xN_y)^2} 
 \sum_{m_x=0}^{N_x-1} \sum_{m_y=0}^{N_y-1}  \sum_{m'_x=0}^{N_x-1} \sum_{m'_y=0}^{N_y-1} 
{\rm Cov}_{\rm sp} [ S_{m_x,m_y}, S_{m'_x, m'_y} ]
e^{-2\pi i[ j_x(m_x-m'_x)/N_x+j_y(m_y-m'_y)/N_y ]}
\nonumber \\ &=&
 \frac{\bar S^2}{(N_xN_y)^2A} 
 \int_{{\mathbb R}^2} d\Delta X\,d\Delta Y\, T(\Delta X,\Delta Y) 
\left| \tilde R\left( \frac{{\bf M}(\Delta X,\Delta Y)}{\lambda z_0} \right) \right|^2
\nonumber \\ && \times
 \sum_{m_x=0}^{N_x-1} \sum_{m_y=0}^{N_y-1}  \sum_{m'_x=0}^{N_x-1} \sum_{m'_y=0}^{N_y-1} 
e^{2\pi i[(M_{11}\Delta X+M_{12}\Delta Y) P(m'_x-m_x) + (M_{21}\Delta X+M_{22}\Delta Y) P(m'_y-m_y)]/(\lambda z_0)}
\nonumber \\ && \times
e^{-2\pi i[ j_x(m_x-m'_x)/N_x+j_y(m_y-m'_y)/N_y ]}.~~~~
\end{eqnarray}
The sums over $m_x$, $m_y$, $m'_x$, and $m'_y$ can be factored, and each one is a geometric series. Using the identity
\begin{equation}
\sum_{j=0}^{N-1} e^{i\alpha j} = e^{i(N-1)\alpha/2}\frac{ \sin (N\alpha/2)}{\sin (\alpha /2)},
\end{equation}
we arrive at
\begin{eqnarray}
\langle |\tilde S_{j_x,j_y}|^2 \rangle_{\rm sp}&=&
 \frac{\bar S^2}{(N_xN_y)^2A} 
 \int_{{\mathbb R}^2} d\Delta X\,d\Delta Y\, T(\Delta X,\Delta Y) 
\left| \tilde R\left( \frac{{\bf M}(\Delta X,\Delta Y)}{\lambda z_0} \right) \right|^2
\nonumber \\ && \times
\left| \frac{\sin \left\{ \pi[j_x + N_xP (M_{11}\Delta X+M_{12}\Delta Y)/(\lambda z_0)] \right\}}{\sin \left\{ \pi[j_x + N_xP (M_{11}\Delta X+M_{12}\Delta Y)/(\lambda z_0)]/N_x \right\}} \right|^2
\nonumber \\ && \times
\left| \frac{\sin \left\{ \pi[j_y + N_yP (M_{21}\Delta X+M_{22}\Delta Y)/(\lambda z_0)] \right\}}{\sin \left\{ \pi[j_y + N_yP (M_{21}\Delta X+M_{22}\Delta Y)/(\lambda z_0)]/N_y \right\}} \right|^2.
\label{eq:PSD-1}
\end{eqnarray}

Equation~(\ref{eq:PSD-1}) accounts exactly for both aliasing of Fourier modes and wavenumber uncertainty from doing the discrete Fourier transform on a finite patch. In the limit that $N_x$ and $N_y$ become large, we can use the rule that the square of the ratio of sines becomes very strongly peaked every time the sine in the denominator is an integer multiple of $\pi$, i.e., it approaches the $\Sha$-function:
\begin{equation}
\left| \frac{\sin (\pi N\zeta)}{\sin (\pi\zeta)} \right|^2 \rightarrow N\Sha(\zeta)
\equiv N\sum_{q=-\infty}^\infty \delta(q-\zeta)
.
\end{equation}
In the context of Eq.~(\ref{eq:PSD-1}), this is
\begin{equation}
\left| \frac{\sin \left\{ \pi[j_x + N_xP (M_{11}\Delta X+M_{12}\Delta Y)/(\lambda z_0)] \right\}}{\sin \left\{ \pi[j_x + N_xP (M_{11}\Delta X+M_{12}\Delta Y)/(\lambda z_0)]/N_x \right\}} \right|^2
\rightarrow N_x \sum_{q_x=-\infty}^\infty \delta\left( \frac{j_x}{N_x} + \frac{P(M_{11}\Delta X+M_{12}\Delta Y)}{\lambda z_0} -q_x \right),
\end{equation}
and similarly in the $y$-direction.
This approximation is valid if $N_x$ and $N_y$ are large and $T$ and $\tilde R$ are continuous functions. The $\delta$-functions in Eq.~(\ref{eq:PSD-1}) collapse the integral over $(\Delta X,\Delta Y)$, albeit with a Jacobian factor. Then Eq.~(\ref{eq:PSD-1}) reduces to
\begin{eqnarray}
\langle |\tilde S_{j_x,j_y}|^2 \rangle_{\rm sp} &\approx &
 \frac{\bar S^2}{N_xN_yA}  \left( \frac{\lambda z_0}P\right)^2 \frac1{\det{\bf M}} \sum_{q_x=-\infty}^\infty \sum_{q_y=-\infty}^\infty
T\left( \frac{\lambda z_0}{P} {\bf M}^{-1} \left(q_x-\frac{j_x}{N_x}, q_y-\frac{j_y}{N_y}\right) \right)
\nonumber \\ && \times
\left| \tilde R \left(\frac{(q_x-j_x/N_x, q_y-j_y/N_y)}{P} \right)  \right|^2.
\label{eq:PSD-2}
\end{eqnarray}
One can see that the PSD is related to the autocorrelation of the aperture ($T$); the absolute value of the Fourier transform of the pixel response function (i.e., pixel contribution to the MTF); and some prefactors. The summation indicates that one should include aliased power as well (as written it is an infinite sum, but in practice usually only a few of the aliased Fourier modes contribute). The factor of $\bar S^2$ is due to the variance of the irradiance equaling the square of the mean (this is because the variance of a $\chi^2$ distribution with 2 degrees of freedom -- a real part and an imaginary part, in this case -- is equal to the square of the mean). The combination $(\lambda z_0)^2/(A\det{\bf M})$ is the coherence area of the speckles in the detector plane (the coherence scale in solid angle is $\lambda^2/A$, and the projection onto the detector gives the factor of $z_0^2$ and the determinant of the distortion matrix), and division by $P^2$ converts it to a number of pixels. The normalization of the PSD at low spatial frequencies, where $T(0)\rightarrow 1$, is set by this coherence area.

Since ${\bf M}$ has eigenvalues that are $<1$ as one moves away from the point on the detector directly below the aperture, a double slit-type aperture (where $T$ consists of triangle functions at a specific separation) probes the MTF at a spatial frequency $q_x-j_x/N_x$ that decreases as one moves away from the center.

A further simplification of Eq.~(\ref{eq:PSD-2}) occurs if one has an aperture and a detector MTF that are simple products of a function in $x$ and a function in $y$, e.g., $T(\Delta X,\Delta Y) = T_X(\Delta X)T_Y(\Delta Y)$, and ${\bf M}$ is close to diagonal ($xy/ z_0^2\ll 1$; in the DCL setup the maximum is 0.026 at the corners of the array). The detector MTF due to the top-hat and Gaussian charge diffusion with rms per axis $\sigma$ would be of this type:
\begin{equation}
\tilde R(u,v) ={\rm sinc} \left( Pu \right) e^{-2\pi^2\sigma^2u^2} \,{\rm sinc}\left( Pv \right)e^{-2\pi^2\sigma^2v^2}.
\end{equation}
(The IPC contribution can't be factored exactly in this way since the diagonal IPC $\alpha_{\rm D}\neq \alpha_{\rm H}\alpha_{\rm V}$, but the diagonal IPC may be small enough that this isn't a big problem for a first investigation.) In this case, the PSD is a product of a PSD in $j_x$ and a PSD in $j_y$, and the PSD can simply be summed over $j_y$:
\begin{equation}
\sum_{j_y=1}^{N_y-1} \langle |\tilde S_{j_x,j_y}|^2 \rangle_{\rm sp} \propto 
 \sum_{q_x=-\infty}^\infty
T_X\left( \frac{\lambda z_0 (q_x-j_x/N_x)}{PM_{11}} \right)
\left| \tilde R_x \left(\frac{q_x-j_x/N_x}{P} \right)  \right|^2.
\label{eq:PSD-3}
\end{equation}

Equation~(\ref{eq:PSD-3}) is the form used for fitting in this paper. Depending on the version of the fit, we either treated the position and width of the peaks in $T_X$ as free parameters, or computed them from the description of the experiment (``forced photometry'' method).

\section{Drift-diffusion model for the modulation transfer function}
\label{appendix_B}

In this appendix, we model the modulation transfer function using a drift-diffusion equation. This is a generalization of the pure diffusion equation that leads to a hyperbolic secant profile \citep{Holloway1986}, which has been used in previous IR detector studies \citep{Barron2007}. 
It also extends the drift-dominated solution, which leads to a Gaussian profile as applied in previous Roman studies \citep{Mosby2020}.
Combinations of drift and diffusion equations have also been applied to charge-coupled device (CCD) detectors in fully and non-fully depleted regimes \citep{Fairfield2006}, although in that case the electric field geometry is different.

We look to model the evolution of the number density of holes within the layer of semiconductor in the Roman detectors, in order to understand the effects of charge diffusion. The change in number density ($n$; units: m$^{-3}$) is dependent on a source ($S$; units: m$^{-3}$ s$^{-1}$) and a flux (${\boldsymbol F}$; units: m$^{-2}$ s$^{-1}$), which can be written as 
\begin{equation}
    \frac{\partial n}{\partial t} = - {\boldsymbol\nabla} \cdot {\boldsymbol F} + S.
\end{equation}
The flux of the holes can be written in terms of the mobility and diffusion coefficients. The mobility term, $\mu$, causes the holes move from the top to the bottom of the detector as a result of the electric field. The diffusion coefficient, $D$, is a result of the random walk the holes will take as they move from high to low densities. These are related via the Einstein relation $\mu = qD/kT$, where $q$ is the hole charge, $k$ is Boltzmann's constant, and $T$ is the temperature. This flux can be written as 
\begin{equation}
    {\boldsymbol F} = -\mu n {\boldsymbol \nabla}\Phi - D{\boldsymbol \nabla} n,
    \label{flux}
\end{equation}
where $\Phi$ is the electric potential.

The charge diffusion timescale is negligible compared with the duration of the exposure, so 
we can set the change in number density equal to 0. We plug in the equation of the flux and take the 2D Fourier transform of the equation (in the $x$ and $y$ directions; we do not Fourier transform $z$),
\begin{equation}
\tilde n(u,v,z) = \int_{{\mathbb R}^2} n(x,y,z)\,e^{-2\pi i (ux+vy)}\,dx\,dy ~~~\Leftrightarrow~~~
n(x,y,z) = \int_{{\mathbb R}^2} \tilde n(u,v,z)\,e^{2\pi i (ux+vy)}\,du\,dv.
\end{equation}
Then we have a differential equation for the number density of holes in the detector:
\begin{equation}
    0 = \frac{\partial }{\partial z}\left [ \mu \frac{d\Phi}{dz}\tilde{n} + D \frac{\partial \tilde{n}}{\partial z}\right] - 4 \pi^{2}\left( u^2 + v^2\right)D \tilde{n} + \tilde{S}.
\label{eq:DE1}
\end{equation}

We solve this differential equation using a Green's function method, setting the source in Fourier space equal to a delta function around $z_0$: $ \tilde{S} = \delta (z - z_0)$. This simplifies the equation where $ z \neq z_0$,  $\tilde{S}(z) = 0$. This simplification leads to a simpler equation: 
\begin{equation}
    0 = \frac{\partial }{\partial z}\left [ \mu \frac{d\Phi}{dz}\tilde{n} + D \frac{\partial \tilde{n}}{\partial z}\right] - 4 \pi^{2}\left( u^2 + v^2\right)D \tilde{n}.
\end{equation}

\begin{figure}
\centering{\includegraphics[width=5in]{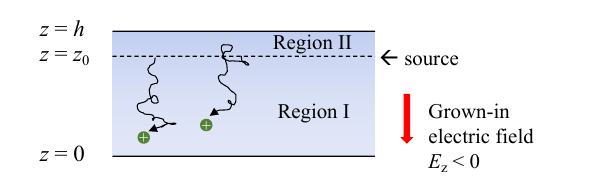}}
\caption{\label{fig:regiondiagram}The solution domain of the differential equation, Eq.~(\ref{eq:DE1}). The illuminated surface (and anti-reflection coating) are at top, and the $p-n$ junctions at which charge is collected are at the bottom.}
\end{figure}

We solve this in the limit of constant electric field $E_z = -d\Phi/dz$. We get two solutions for the number density in the two regions I and II (see Fig.~\ref{fig:regiondiagram}; we indicate these with ``I'' or ``II'' subscripts):

\begin{equation}
   \tilde{n}_{\rm I} =  c_{1\,\rm I}e^{r_{+}z} + c_{2\,\rm I}e^{r_{-}z}
   \label{diff-sol1}
\end{equation}
and
\begin{equation}
     \tilde{n}_{\rm II} = c_{1\,\rm II}e^{r_{+}z} + c_{2\,\rm II}e^{r_{-}z},
     \label{diff-sol2}
\end{equation}    
where 
\begin{equation}
    r_{\pm} = \frac{\mu E_z \pm \sqrt{\mu^2 (-E_z)^2 + 16D^2\pi^2(u^2 + v^2)}}{2D} .
\end{equation}

The junction conditions at $z_0$ are the continuity of $\tilde n$ and the step function in $\partial\tilde n/\partial z$ obtained by integrating from $z_0-\epsilon$ to $z_0+\epsilon$:
\begin{equation}
    \tilde{n}(z_0 + \epsilon) = \tilde{n}(z_0 - \epsilon)
    \label{junction1}
\end{equation}
and
\begin{equation}
    0 = D \left[\frac{\partial \tilde{n}}{\partial z}(z_0 + \epsilon) - \frac{\partial \tilde{n}}{\partial z}(z_0 - \epsilon)\right] + 1.
    \label{junction2}
\end{equation}
The boundary conditions are found at the top, where $z=h$, and bottom, where $z=0$, of the detector. At the bottom of the detector, we expect there to be no holes as we expect all charges to have been collected. At the top of the detector, we expect the flux to be 0, as charges should not be moving through this boundary. The equations for these conditions are
\begin{equation}
    \tilde{n}(0) = 0 
    \label{boundary1}
\end{equation}
and
\begin{equation}
    \tilde{F}_z(h) = 0 = \mu E_z \tilde{n}(h) - D\frac{\partial \tilde{n}}{\partial z}(h).
    \label{boundary2}
\end{equation}

With four unknown constants and four boundary and junction conditions, we can solve for each of the constants in terms of one another.

Starting in region I, using boundary condition Eq.~(\ref{boundary1}) in Eq.~(\ref{diff-sol1}), we find a relationship between the coefficients: $c_{1\,\rm I} = -c_{2\,\rm I}$, allowing us to rewrite the number density equation in region I as
\begin{equation}
    \tilde{n}_{\rm I} = c_{1\,\rm I}(e^{r_+z} - e^{r_-z})
\end{equation}  
In region II, using boundary condition Eq. (\ref{boundary2}) in Eq.(\ref{diff-sol2}), we get the relation
\begin{equation}
    c_{2\,\rm II} = -c_{1\,\rm II} e^{h(r_+ - r_-)}\left(\frac{Dr_+ - \mu E_z}{Dr_- - \mu E_z}\right)
~~~\Rightarrow~~~
    c_{2\,\rm II} = -c_{1\,\rm II} e^{h(r_+ - r_-)}\left(\frac{r_-}{r_+}\right).
\end{equation}

Using the first junction condition (Eq.~\ref{junction1}) and rewriting so the equation is in terms of $c_{1\,\rm II}$,  
\begin{equation}
    c_{1\,\rm II} = c_{1\,\rm I} \frac{r_+(e^{r_+z_0}-e^{r_-z_0})}{r_+e^{r_+z_0} - r_-e^{r_-z_0}e^{h(r_+ - r_-)}}.
\end{equation}

Using the second junction condition Eq. (\ref{junction2}) and similarly rewriting the equation in terms of $c_{1I}$, 
we may solve for $c_{1\,\rm I}$:
\begin{equation}
    c_{1\,\rm I} = -\frac{1}{D}\left[\frac{(e^{r_+z_0}-e^{r_-z_0})(r_-^2e^{r_-z_0}e^{h(r_+-r_-)} - r_+^2e^{r_+z_0})}{r_+e^{r_+z_0} - r_-e^{r_-z_0}e^{h(r_+-r_-)}} + r_+e^{r_+z_0} - r_-e^{r_-z_0}\right]^{-1}.
\end{equation}

We are interested the flux at the bottom of the detector, $\tilde F_z(0)$, as the MTF is the negative of this value (for unit source and recalling that collected charges move toward negative $z$). We recall Eq.~(\ref{flux}) now at $z=0$,  
\begin{equation}
    \tilde F_z(0) = \mu\tilde{n}(0)E_z - D \frac{\partial\tilde{n}}{\partial z}(0)
    = -Dc_{1\,\rm I}(r_+ - r_-).
\end{equation}

Plugging in the value of $c_{1\,\rm I}$ into this equation for the flux, we get
\begin{eqnarray} 
\tilde F_z(0) = \left[\frac{(e^{r_+z_0}-e^{r_-z_0})(r_-^2e^{r_-z_0}e^{h(r_+-r_-)} - r_+^2e^{r_+z_0})}{r_+e^{r_+z_0} - r_-e^{r_-z_0}e^{h(r_+-r_-)}} + r_+e^{r_+z_0} - r_-e^{r_-z_0}\right]^{-1}(r_+ - r_-).
\end{eqnarray}
as the solution of the Green's equation. Using the negative of this solution and the absorption probability over the detector, we average over $z_0$ to find the charge diffusion contribution to the MTF of the Nancy Grace Roman detectors. 

\section{Alternative Analysis}
\label{appendix_C}

\begin{figure}[H]
  \centering
  \begin{minipage}{0.48\textwidth}
    \centering    \includegraphics[width=\linewidth]{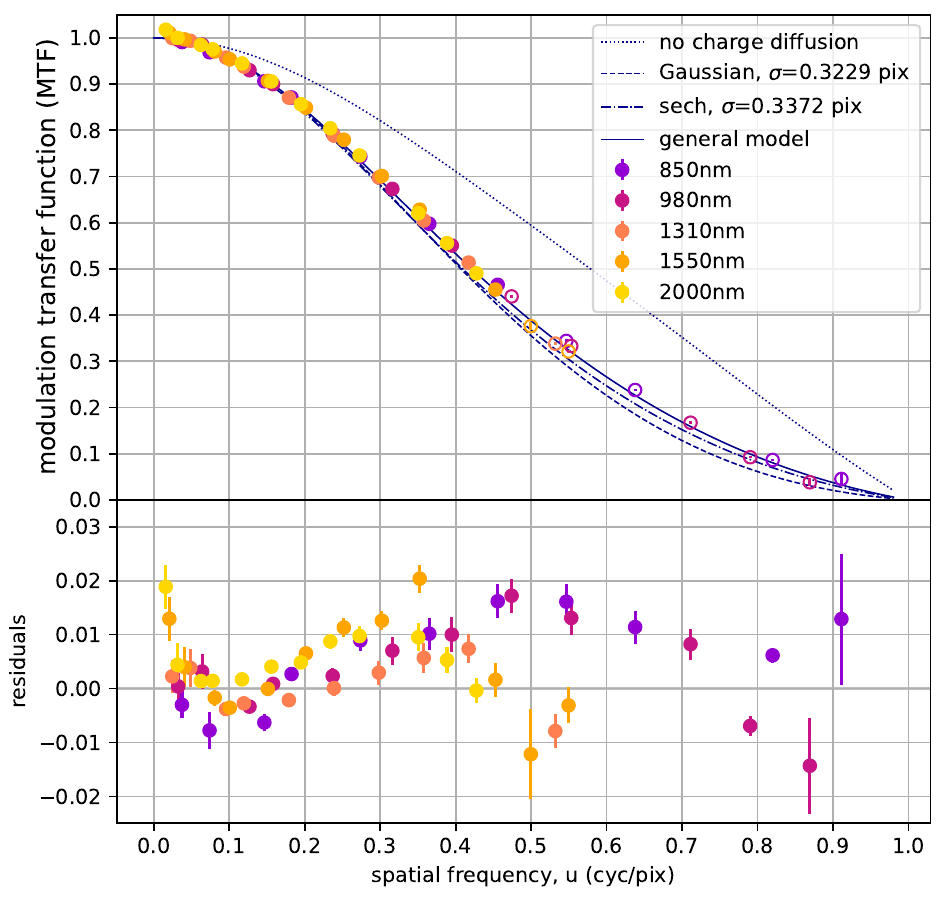}
    \caption{MTF data from frames 1-3 used to fit the general charge diffusion model, likely including nonlinear effects not present in frames 1-2. The closed points were found using the original fit, while the open points are the forced fit to the spatial frequencies. The MTF model without charge diffusion and the special cases for charge diffusion are compared. The residuals of the robust model are plotted at the bottom, indicating periodic behavior.}
    \label{fig:F3}
  \end{minipage}\hfill
  \begin{minipage}{0.48\textwidth}
    \centering \includegraphics[width=\linewidth]{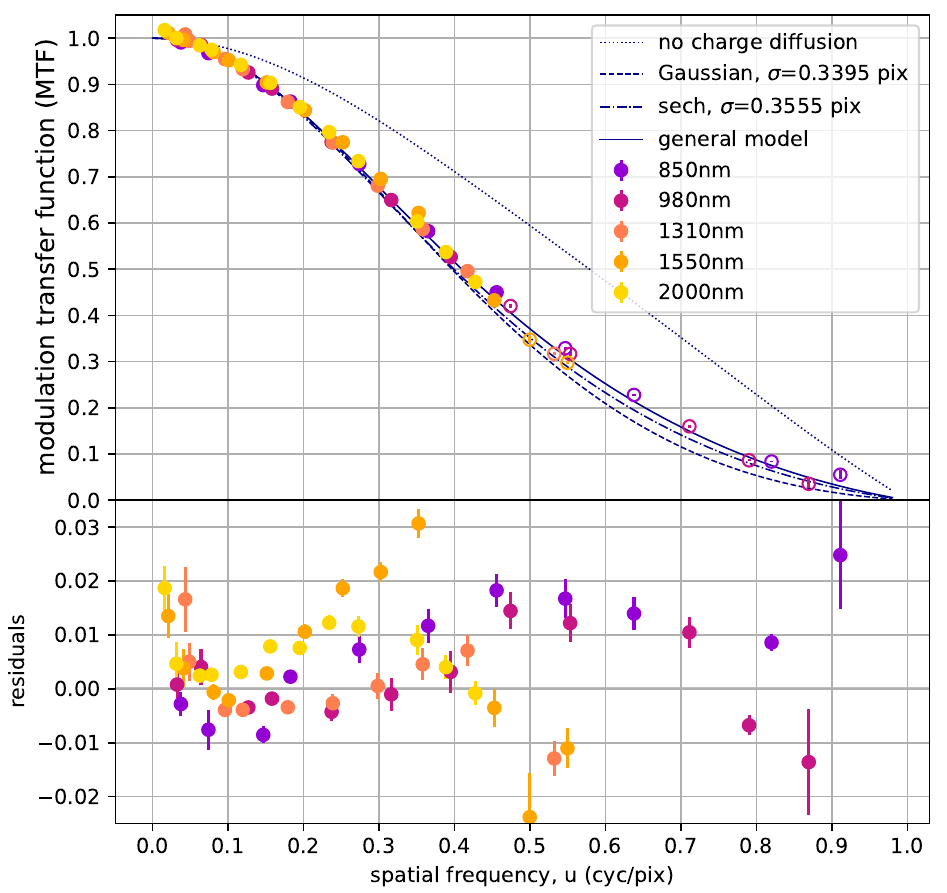}
    \caption{MTF data from frames 1-5 used to fit the general model, likely including the most significant nonlinear effects. The closed points were found using the original fit, while the open points are the forced fit to the spatial frequencies. The MTF model without charge diffusion and the special case models are displayed. The residuals of the robust model are plotted at the bottom, indicating periodic behavior with clear contributions from nonlinear effects.}
    \label{fig:F5}
  \end{minipage}
\end{figure}

The charge diffusion model was also produced using the MTF data from F1-3 and F1-5 to improve the accuracy of the model. These models show the results of increasing nonlinearities in the data which are introduced over time in Fig. \ref{fig:F3} and Fig. \ref{fig:F5}. Each model appears to follow similarly to Fig. \ref{fig:F2-full}, but present more drastic residuals. Both display the same slightly periodic behavior with significant discrepancies between the high and low wavelengths at u = 0.5 cy/pix. Further discrepancies appear around u = 0.25 cy/pix and 0.9 cy/pix. The source of these discrepancies are not yet understood.

\begin{figure}
    \centering
    \includegraphics[width=0.5\linewidth]{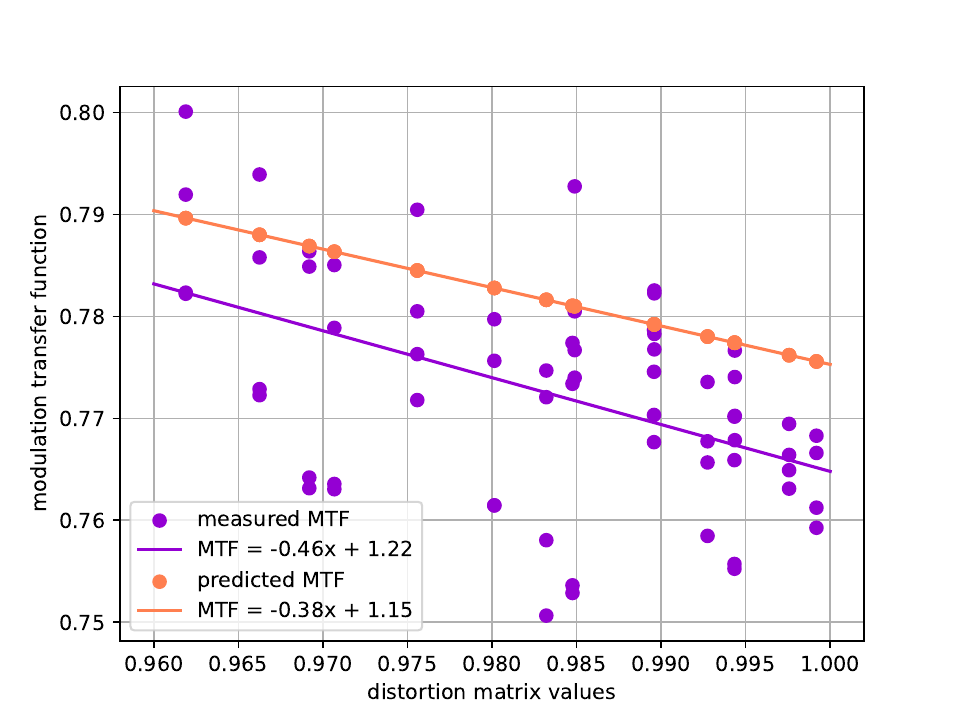}
    \caption{Measured and predicted MTF for aperture 7 in the 1550 nm in 64 regions. The values are plotted with respect to the values of the distortion matrix to indicate how the MTF varies with spatial frequency. The predicted MTF has a slope of -0.377 $\pm$ 0.000095, with an intercept of 1.15 $\pm$ 0.000094. The slope of the measured MTF is -0.46 $\pm$0.10 with an intercept of 1.22 $\pm$ 0.10. The measured slope and intercept is consistent with the predicted values within 1$\sigma$.}
    \label{fig:distortioncompare}
\end{figure}

The introduction of the non-small angle effects of the double slit were introduced to the prediction of the spatial frequencies for the forced fit. To determine how the MTF varies across the detector, we compare the predicted MTF, according to the general model, to the values from the fitting procedure. The measured values of the MTF are from aperture 7 in 1550 nm. These were chosen for this check because the SNR was high, with each point from the 64 regions displayed in Fig. \ref{fig:datavis}. This comparison is shown in Fig. \ref{fig:distortioncompare}, where both values of MTF are plotted against the distortion matrix values. The slope of the measured MTF is within 1$\sigma$ of the predicted slope, with a MTF=0.1 difference in the intercepts, also within 1$\sigma$. The parallel slopes indicate the MTF consistently varies across the detector as predicted, and the intercept indicates that there is a systematic offset in the measurement of the MTF that results in a lower value than expected.

\bibliography{sample631}{}
\bibliographystyle{aasjournal}

\end{document}